\documentclass[aps,preprint,superscriptaddress,groupedaddress,nofootinbib]{revtex4}

\usepackage{graphicx}
\usepackage[normalem]{ulem}
\usepackage[export]{adjustbox}
\usepackage{float}
\usepackage{amsmath}
\usepackage{amssymb}
\usepackage[rightcaption]{sidecap}
\usepackage{hyperref}
\usepackage{xcolor}

\usepackage{dcolumn}
\usepackage{xcolor}
\usepackage{slashed}

\usepackage{multirow}

\graphicspath{ {./plots/} }
\def\beq{\begin{equation}}
\def\eeq{\end{equation}}
\def\bea{\begin{eqnarray}}
\def\eea{\end{eqnarray}}

\def\figureautorefname~#1\null{Fig.\,#1\null}
\def\tableautorefname~#1\null{Tab.\,#1\null}

\def\equationautorefname~#1\null{Eq.\,(#1)\null}
\allowdisplaybreaks[4]

\makeatletter

\begin{document}

\title{Resonant top pair searches at the LHC:\\ a window to electroweak phase transition}

\author{Dorival Gon\c{c}alves}
\email{dorival@okstate.edu}
\affiliation{Department of Physics, Oklahoma State University, Stillwater, OK, 74078, USA}
\author{Ajay Kaladharan}
\email{kaladharan.ajay@okstate.edu}
\affiliation{Department of Physics, Oklahoma State University, Stillwater, OK, 74078, USA}
\author{Yongcheng Wu}
\email{ycwu@njnu.edu.cn}
\affiliation{Department of Physics and Institute of Theoretical Physics, Nanjing Normal University, Nanjing, 210023, China}
\affiliation{Department of Physics, Oklahoma State University, Stillwater, OK, 74078, USA}

\begin{abstract}
The dynamics of electroweak phase transition could have profound consequences for particle physics and cosmology. We study the prospects for the HL-LHC to probe the strong first-order electroweak phase transition (SFOEWPT) regime in the type-I 2HDM. We focus on the Higgstrahlung channel $pp\to ZH/A$ with a resonant top-quark pair final state $H/A\to t\bar t$. We find that the top-quark pair final state renders the largest sensitivity to the SFOEWPT regime, in comparison to the other Higgstrahlung searches already performed by ATLAS and CMS, that focus on the $H/A\to bb$ and $H\to WW$ final states. We also derive the complementarity of the Higgstrahlung searches with other relevant classes of searches at the HL-LHC and compare them with the gravitational wave sensitivity at LISA.
\end{abstract}

\maketitle
\flushbottom
\clearpage

\section{Introduction}
\label{sec:intro}

The electroweak symmetry appears exact in the early Universe at high temperatures. However, at temperatures around 100~GeV, the Higgs field develops a vacuum expectation value, spontaneously breaking this symmetry. At this time, the Universe goes through a transition from a symmetric to a broken phase. While the Standard Model (SM) predicts a continuous transition~\cite{Kajantie:1996mn}, new physics can generate a first-order phase transition. As yet, it is unknown how the electroweak phase transition (EWPT) occurred -- whether it was violent or calm; first-order phase transition or smooth crossover.

First-order phase transition is a violent phenomenon that can display relevant consequences for the evolution of the Universe. In particular, this transition could provide the required out-of-equilibrium conditions to generate the  baryon asymmetry of the Universe via electroweak baryogenesis~\cite{Sakharov:1967dj,Trodden:1998ym,Cohen:1993nk,Carena:1996wj,Morrissey:2012db}. If the first-order phase transition was strong enough, it could have generated relic gravitational waves (GW) that might be probed in future GW detectors, such as in the space-based LISA mission~\cite{Grojean:2006bp,Audley:2017drz}. This transfiguration in the EWPT profile usually requires novel degrees of freedom close to the electroweak scale, displaying sizable interactions with the Higgs boson~\cite{Ramsey-Musolf:2019lsf,Goncalves:2021egx}. Thus, precision measurements for the Higgs sector and the search for new heavy scalars at colliders are also important avenues for probing models which trigger first-order electroweak phase transition. The Higgs pair production $pp\to hh$ plays a central role in these studies, as it provides a direct probe to the Higgs potential through resonant and non-resonant searches~\cite{Eboli:1987dy,Plehn:1996wb,No:2013wsa,Goncalves:2018qas,Barman:2020ulr}. Thus, this subject provides an exciting research arena on the interface between particle physics and cosmology.

The strength of the phase transition is correlated with the potential upliftment of the true vacuum compared to the symmetric one at zero temperature~\cite{Dorsch:2017nza,EWPT-NMSSM,EWPT-Nature}. This gauge-invariant property sheds light on the phase transition pattern analytically. In particular, it is possible to derive that the large order parameter, $\xi_c\equiv v_c/T_c\gtrsim 1$, favors light scalar masses~\cite{Goncalves:2021egx}. This grants extra motivation for resonant searches at the Large Hadron Collider (LHC).  Besides scalar masses under the TeV scale, the analytical structure of the beyond the SM effects on the vacuum upliftment, leading to strong first-order electroweak phase transition (SFOEWPT), $\xi_c\gtrsim 1$, results in a distinct hierarchy of masses among the new scalar states.

In the two Higgs doublet model (2HDM)~\cite{Dorsch:2013wja,Basler:2016obg,Dorsch:2016tab,Bernon:2017jgv,Dorsch:2017nza,Andersen:2017ika,Kainulainen:2019kyp,Su:2020pjw,Li:2020hao,Kling:2020hmi,Davoudiasl:2021syn,Biekotter:2021ysx,Aoki:2021oez,Enomoto:2021dkl,Atkinson:2022pcn,Anisha:2022hgv}, the parameter space  $m_H<m_{H^\pm}\approx m_A$, with large mass difference, yields a favorable regime to achieve SFOEWPT~\cite{Dorsch:2014qja,Goncalves:2021egx}. Due to the preference for large mass hierarchy among the scalar modes, it is likely that at least one of the scalar states is above the top-quark pair threshold. As recently shown, the gluon fusion production channel $gg\to A/H\to t\bar t$ plays a leading role in these phenomenological studies, granted by its large event rate~\cite{Goncalves:2021egx}. Another relevant channel is the resonant Higgstrahlung production $A\to ZH$~\cite{Dorsch:2014qja}, which is enhanced in the preferred scalar mass regime for SFOEWPT, $m_H<m_{H^\pm}\approx m_A$. However, the current experimental analyses explore the $A\to ZH$ searches only through the decays $H\to b\bar b$ and $H\to WW$ with $Z\to\ell\ell$~\cite{Aad:2020ncx}. The flipped channel $H\to ZA$ is also analyzed with $A\to b\bar b$ and $Z\to\ell\ell$~\cite{Sirunyan:2019wrn}. For $A\to ZH$ or $H\to ZA$ channels, the corresponding heavy scalar decay to top-quark pair $H/A \to t\bar t$ can also be an important signature for the SFOEWPT parameter space. Compared to the already explored scalar decays to $b\bar b$ and $W^+W^-$, the $t\bar t$ final state covers a different mass spectrum, which can significantly improve the sensitivity of $A\to ZH$ and $H\to ZA$ channels. In the present work, we scrutinize the sensitivity of the scalar decays to top-quark pair for probing SFOEWPT in the 2HDM. Special attention will be devoted to the Higgstrahlung mode $pp\to Z H/A$ at the high luminosity LHC (HL-LHC).

This paper is organized as follows. In~\autoref{sec:theory}, we introduce the two Higgs doublet model. In~\autoref{sec:AZH}, we discuss the leading contributions for the 2HDM signature $pp\to ZH/A$  with top-quark pair final state $H/A\to t\bar t$. The corresponding HL-LHC sensitivity is derived in~\autoref{sec:ana}. In~\autoref{sec:EWPT}, we study the complementarity between collider and gravitational wave experiments to probe the electroweak phase transition profile in the 2HDM. In particular, we scrutinize the relevance of the Higgstrahlung channel with top pair final states $H/A\to t\bar t$ with respect to other relevant classes of searches at the HL-LHC. We also contrast the collider with the gravitational wave sensitivity. We summarize in~\autoref{sec:summary}. Further details on the complementary channel  $gg\to H/A\to t\bar t$ are presented in~\autoref{sec:ggF}.

\section{Two Higgs Doublet Model}
\label{sec:theory}

In this work, we consider the CP-conserving 2HDM with a softly broken $\mathbb{Z}_2$ symmetry, where the scalar potential can be written as~\cite{Branco:2011iw}
\begin{align}
    V(\Phi_1,\Phi_2) =& m_{11}^2\Phi_1^\dagger\Phi_1 + m_{22}^2\Phi_2^\dagger\Phi_2 - m_{12}^2(\Phi_1^\dagger\Phi_2 + \text{h.c.}) + \frac{\lambda_1}{2}(\Phi_1^\dagger\Phi_1)^2 + \frac{\lambda_2}{2}(\Phi_2^\dagger\Phi_2)^2\nonumber \\
            & + \lambda_3(\Phi_1^\dagger\Phi_1)(\Phi_2^\dagger\Phi_2) + \lambda_4(\Phi_1^\dagger\Phi_2)(\Phi_2^\dagger\Phi_1) + \frac{\lambda_5}{2}\left((\Phi_1^\dagger\Phi_2)^2 + \text{h.c.}\right).
\label{equ:v_tree}
\end{align}
Expanding around the VEVs, the two $SU(2)_L$ doublets  can be expressed by
\begin{align}
    \Phi_1 = \left(\begin{array}{c}
        \phi_1^+\\
        \frac{v_1+\phi_1^0+i\eta_1}{\sqrt{2}}
    \end{array}\right),\quad \Phi_2 = \left(\begin{array}{c}
        \phi_2^+\\
        \frac{v_2+\phi_2^0+i\eta_2}{\sqrt{2}}
    \end{array}\right),
\end{align}
where the vacuum expectation values $v_i$ are connected to the SM VEV by $v_1^2+v_2^2=v^2\approx(246~\text{GeV})^2$.
After electroweak symmetry breaking, the model provides five physical mass eigenstates from the scalar sector: two CP-even neutral scalars $h$ and $H$, where $h$ is identified as the SM Higgs boson, a CP-odd neutral scalar $A$, and a pair of charged scalars $H^\pm$. The mass and gauge eigenstates are related by the rotation angle $\beta$ for the charged and CP-odd sectors,  $\tan\beta \equiv v_2/v_1$, and by the angle $\alpha$ for the CP-even sector
\begin{align}
    \left(\begin{array}{c}
        G^\pm\\
        H^\pm
    \end{array}\right) =
    \mathcal{R}(\beta)
    \left(\begin{array}{c}
        \phi_1^\pm\\
        \phi_2^\pm
    \end{array}\right),\,\,\,
    \left(\begin{array}{c}
        G^0\\
        A
    \end{array}\right)= \mathcal{R}(\beta)
    \left(\begin{array}{c}
        \eta_1\\
        \eta_2
    \end{array}\right),\,\,\,
    \left(\begin{array}{c}
        H\\
        h
    \end{array}\right)= \mathcal{R}(\alpha)
    \left(\begin{array}{c}
        \phi_1^0\\
        \phi_2^0
    \end{array}\right).
\end{align}
The rotation matrix is given by
\begin{align}
    \mathcal{R}(x)=\left(\begin{array}{cc}
       c_x  & s_x \\
       -s_x & c_x
   \end{array}\right)\,,
\end{align}
where $s_x \equiv \sin x$ and $c_x \equiv \cos x$. The charged and neutral massless Goldstone bosons are denoted by $G^\pm$ and $G^0$, respectively.

The physical parameters of 2HDM can be chosen as
\begin{align}
    \tan\beta,~\cos(\beta-\alpha),~m_{12}^2,~v,~m_h(=125~{\rm GeV}),~m_H,~m_A,~m_{H^\pm}.
    \label{eq:scan}
\end{align}
The parameters $t_\beta\equiv \tan\beta$ and $c_{\beta-\alpha}\equiv \cos(\beta-\alpha)$ control the coupling strength of the scalar particles to fermions and gauge bosons, displaying critical phenomenological relevance.  Whereas the Higgs-gauge couplings scale as $g_{hVV}\propto s_{\beta-\alpha}$ and $g_{HVV}\propto c_{\beta-\alpha}$, the fermion interactions hinge on both $t_\beta$ and $c_{\beta-\alpha}$. The fermionic couplings depend on the  $\mathbb{Z}_2$ charge assignment in the Yukawa sector. In this study, we focus on the Type-I scenario where all fermions couple solely to $\Phi_2$. When confronted with current experimental constraints, the alignment limit  $c_{\beta-\alpha}\to0$ is preferred~\cite{Gunion:2002zf,Han:2020lta}. Furthermore, electroweak precision measurements also put strong constraints on the 2HDM parameter space, which pushes either $\Delta m_H\equiv m_H-m_{H^\pm}$ or $\Delta m_A\equiv m_A-m_{H^\pm}$ close to zero.

\section{Top Pair Resonant Searches via \texorpdfstring{$pp\to ZH/A$}{pptoZHA}}
\label{sec:AZH}

\begin{figure}
    \centering
    \includegraphics[width=\textwidth]{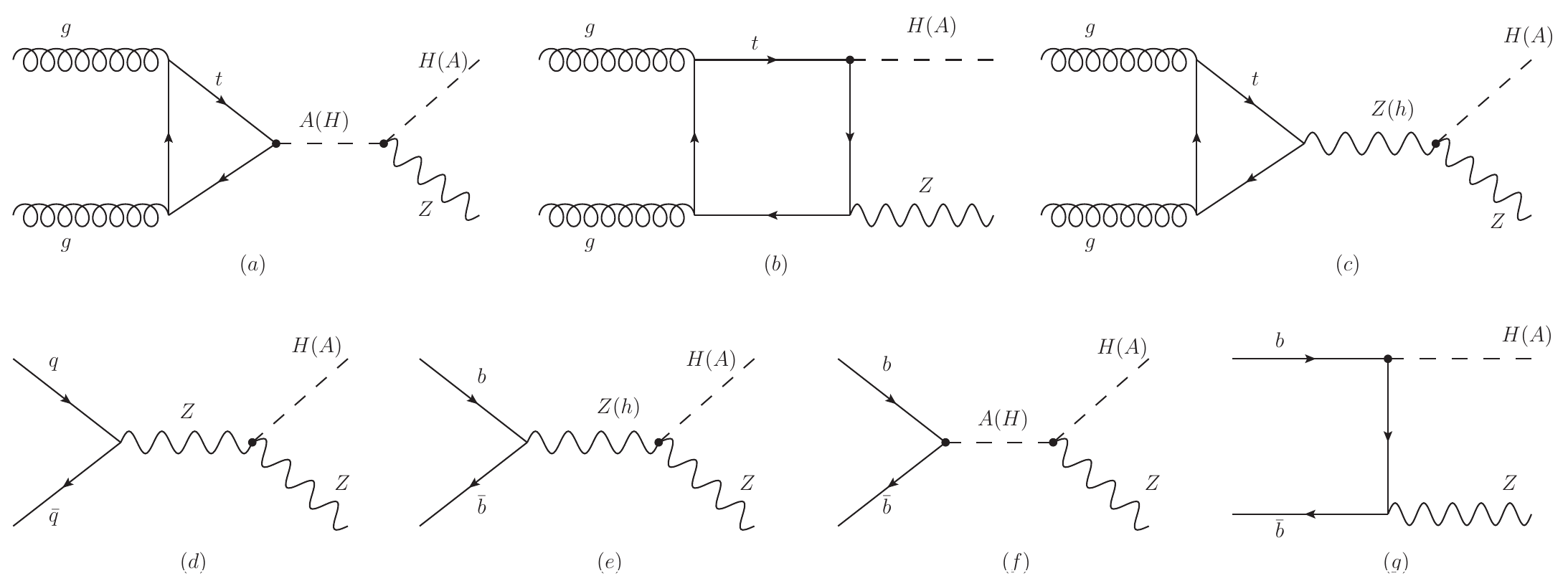}
    \caption{Representative set of Feynman diagrams for Higgstrahlung production $pp\to ZH/A$ at the LHC:
    (a-c)~loop induced gluon fusion production $gg\to ZH/A$; (d)~Drell-Yan like contribution $q\bar q\to ZH$; and (e-g)~bottom quark initiated production $b\bar b\to ZH/A$. We highlight the couplings that differ between the SM and 2HDM.}
    \label{fig:AZH_FeynDiag}
\end{figure}

In this section, we will discuss the dominant contributions for the 2HDM signature $pp\to ZH/A$ with top-quark pair final state $H/A\to t\bar{t}$. The Higgstrahlung channel, $pp\to ZH/A$, has three dominant contributions at the LHC~\cite{Goncalves:2015mfa}. First, it contains the loop-induced gluon fusion production $gg\to ZH/A$. This mode encapsulates both  resonant $gg\to A/H\to ZH/A$  and non-resonant  contributions, see  \autoref{fig:AZH_FeynDiag}~(a-c)~\footnote{Note that we grouped two diagrams with two different types of the propagators in~\autoref{fig:AZH_FeynDiag}~(c), as they share the same scaling behavior as we will discuss in the following. Similar for~\autoref{fig:AZH_FeynDiag}~(e).}. Another relevant mode stems from the tree-level Drell-Yan-like production, $q\bar{q}\to Z^*\to ZH$,~\autoref{fig:AZH_FeynDiag}~(d). Notably, this production channel does not result in  a correspondent $ZA$ final state. This mode is absent at this order of perturbation expansion, as the $ZZA$ coupling is forbidden at tree-level.  Finally, there are also important $b$-quark initiated contributions, $b\bar{b}\to ZH/A$, that can be augmented in the enhanced bottom Yukawa coupling regime,  \autoref{fig:AZH_FeynDiag}~(e-g).

\begin{figure}
    \centering
    \includegraphics[width=0.48\textwidth]{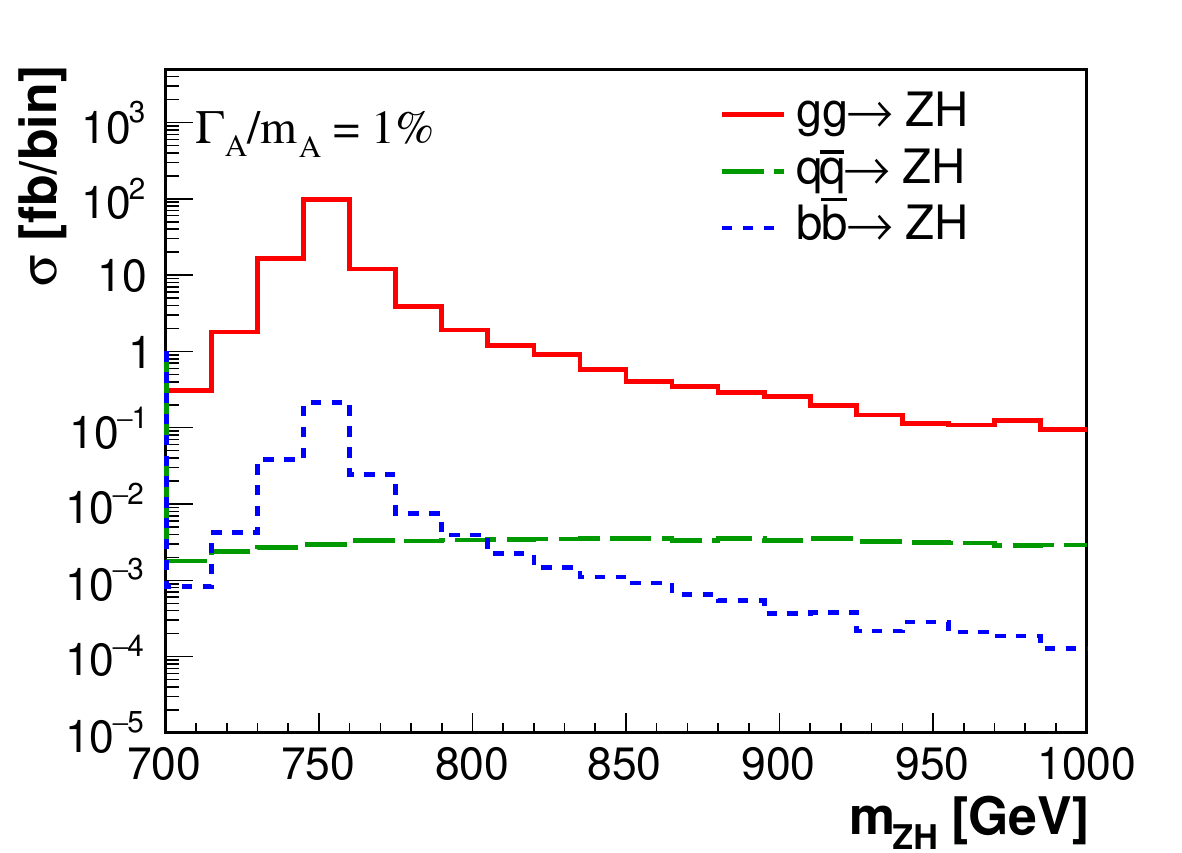}
    \includegraphics[width=0.48\textwidth]{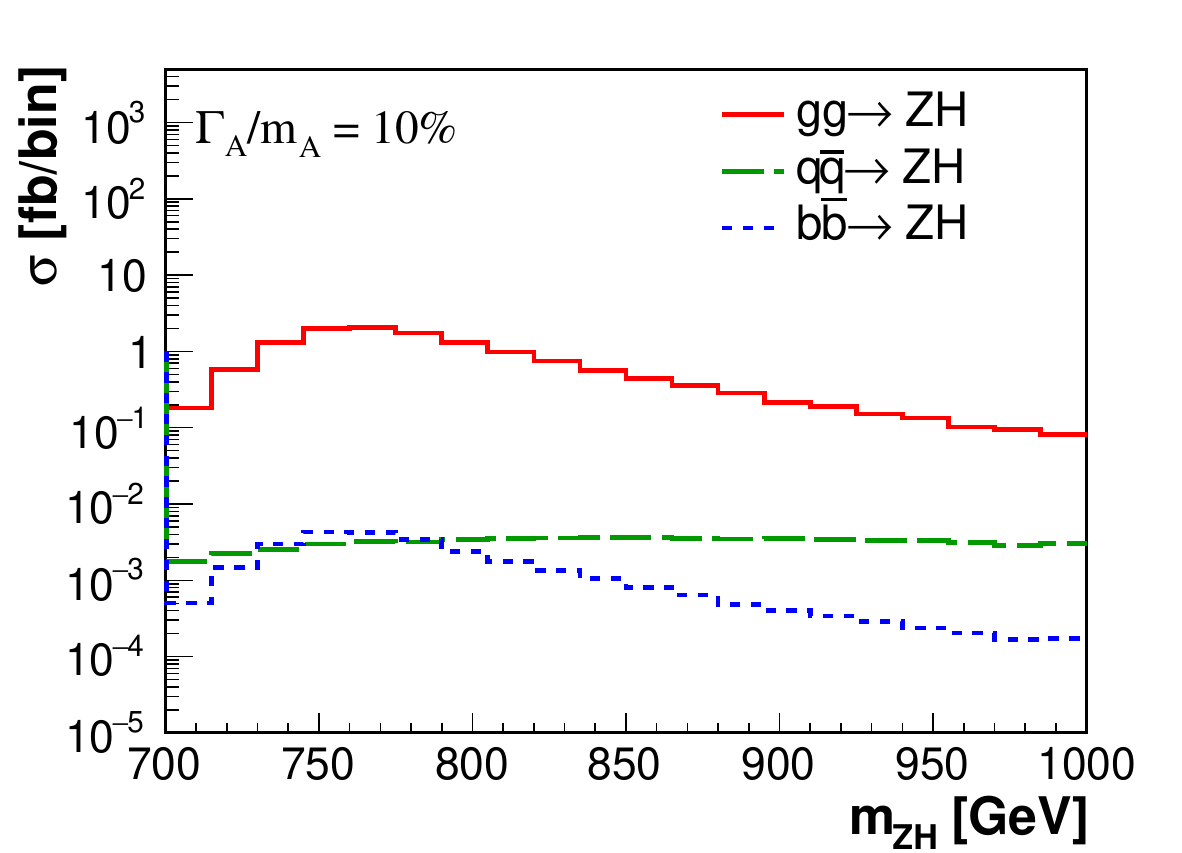}
    \caption{{Invariant mass distribution $m_{ZH}$ for the Higgstrahlung signal $pp\to ZH$ at the parton level with different choice of scalar widths: $\Gamma_A/m_A=1\%$ (left panel) and $\Gamma_A/m_A=10\%$ (right panel). We decompose the signal in its leading contributions:  $i)$~loop-induced gluon fusion production $gg\to ZH$ (red); $ii)$~$b$-quark initiated production $b\bar{b}\to ZH$ (blue); and $iii)$~Drell-Yan like contribution $q\bar{q}\to Z^*\to ZH$ (green). For illustration, we consider the type-I 2HDM with maximally allowed mixing angles, giving the current LHC data: $c_{\beta-\alpha}\approx 0.3$. In addition, we assume $m_H = 600$ GeV, $m_A  = 750$~GeV,  and $t_\beta = 1$.}
    }
    \label{fig:signal-terms}
\end{figure}

In~\autoref{fig:signal-terms}, we illustrate these three leading contributions to the Higgstrahlung signal $pp\to ZH$. The gluon fusion and bottom quark-initiated channels generally result in significant rates, typically enhanced by intermediate resonant Higgs production $H/A\to ZA/H$. Between these two terms, there are three leading effects that guarantee the dominance of the gluon fusion channel: $i)$~the gluon fusion channel is driven by the large parton distribution function of initial state gluons; $ii)$~it has larger initial state color factor; and $iii)$~the triangle and box diagrams are enhanced by the sizable top Yukawa coupling, shown respectively in \autoref{fig:AZH_FeynDiag}~(a) and \autoref{fig:AZH_FeynDiag}~(b). The last term, the Drell-Yan-like mode, generally produces subleading corrections. Its cross-section is hampered by the absence of a resonant scalar mode $A\to ZH$ and the dependence with the mixing angle $\sigma_\text{DY}^{ZH}\propto c_{\beta-\alpha}^2$. The suppressed rate becomes apparent even when adopting maximally allowed mixing angles. We depict this scenario in \autoref{fig:signal-terms}, considering the Type-I 2HDM with $c_{\beta-\alpha}\approx 0.3$. Remarkably, the type-II scenario leads to further depleted Drell-Yan production, as the experimental constraints tend to confine the model parameters further towards the alignment limit, $c_{\beta-\alpha}\to 0$.

\begin{figure}
    \centering
    \includegraphics[width=0.48\textwidth]{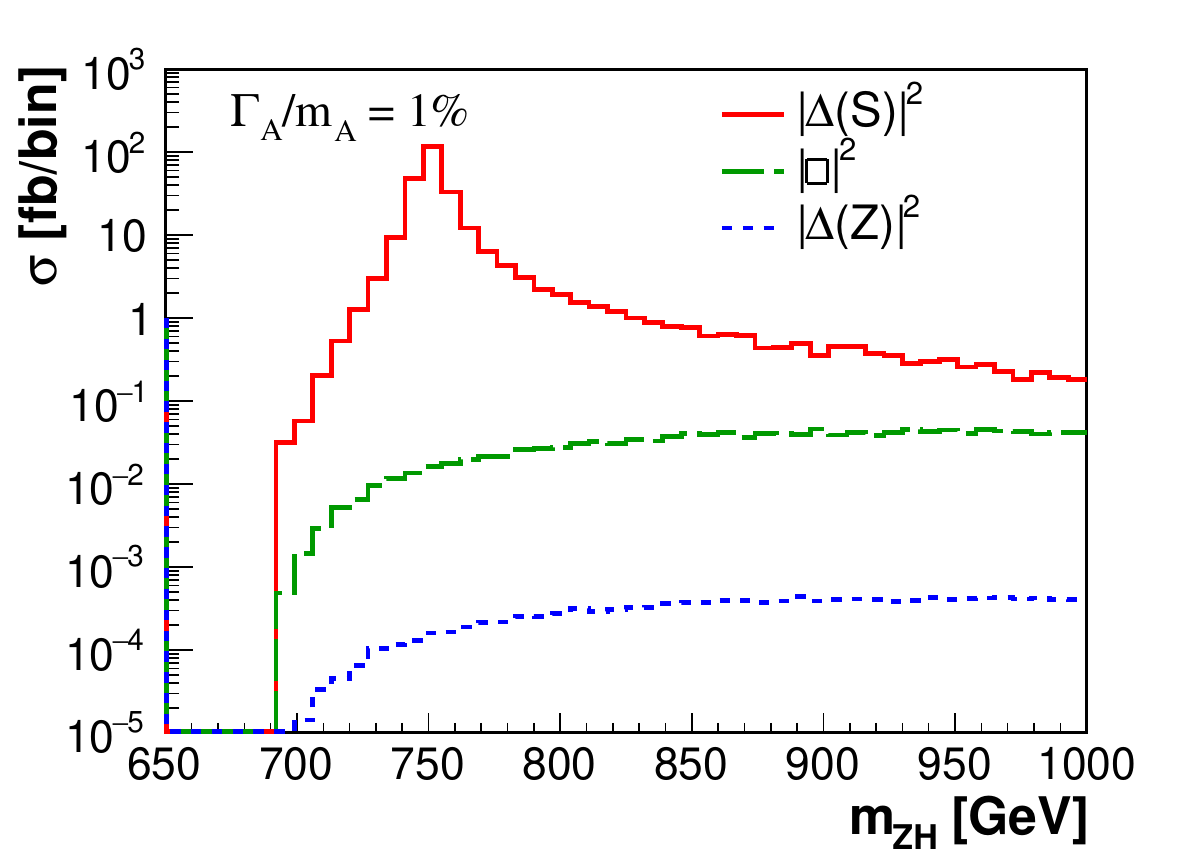}
    \includegraphics[width=0.48\textwidth]{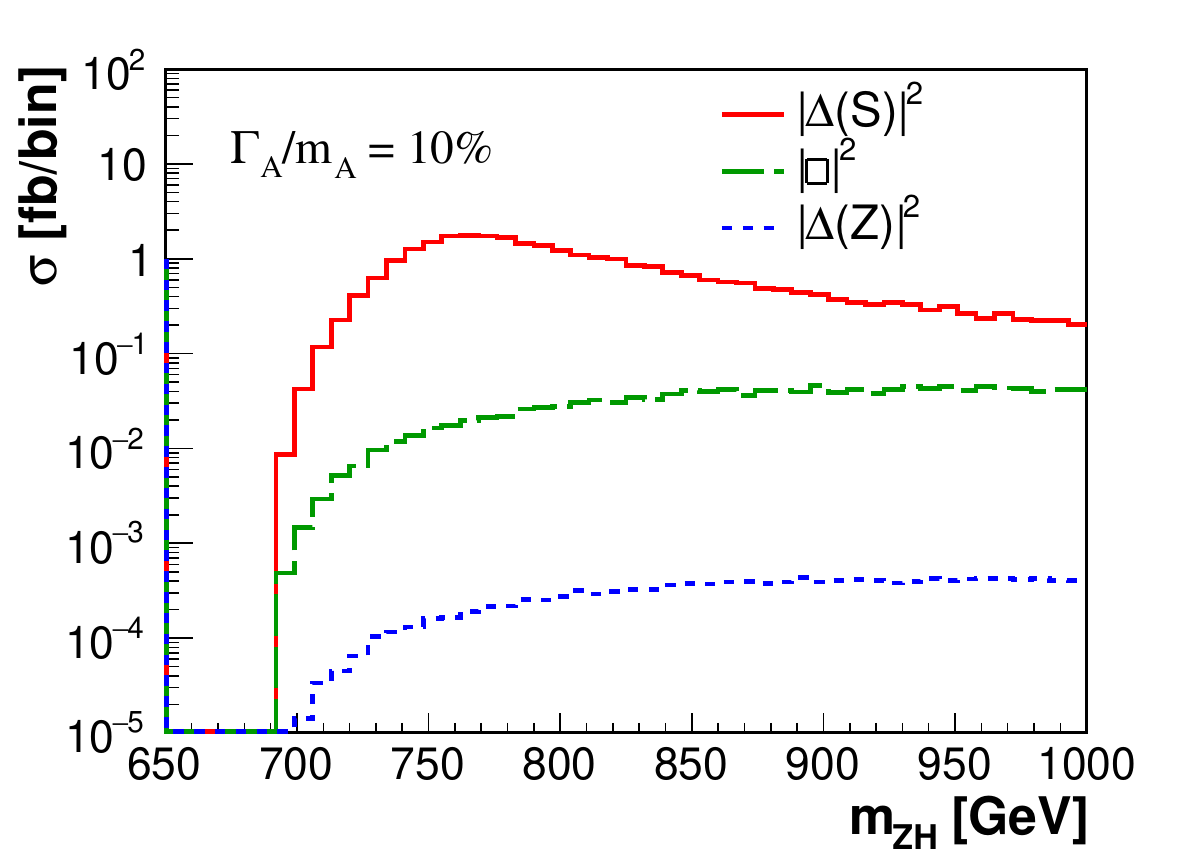}\\
    \includegraphics[width=0.48\textwidth]{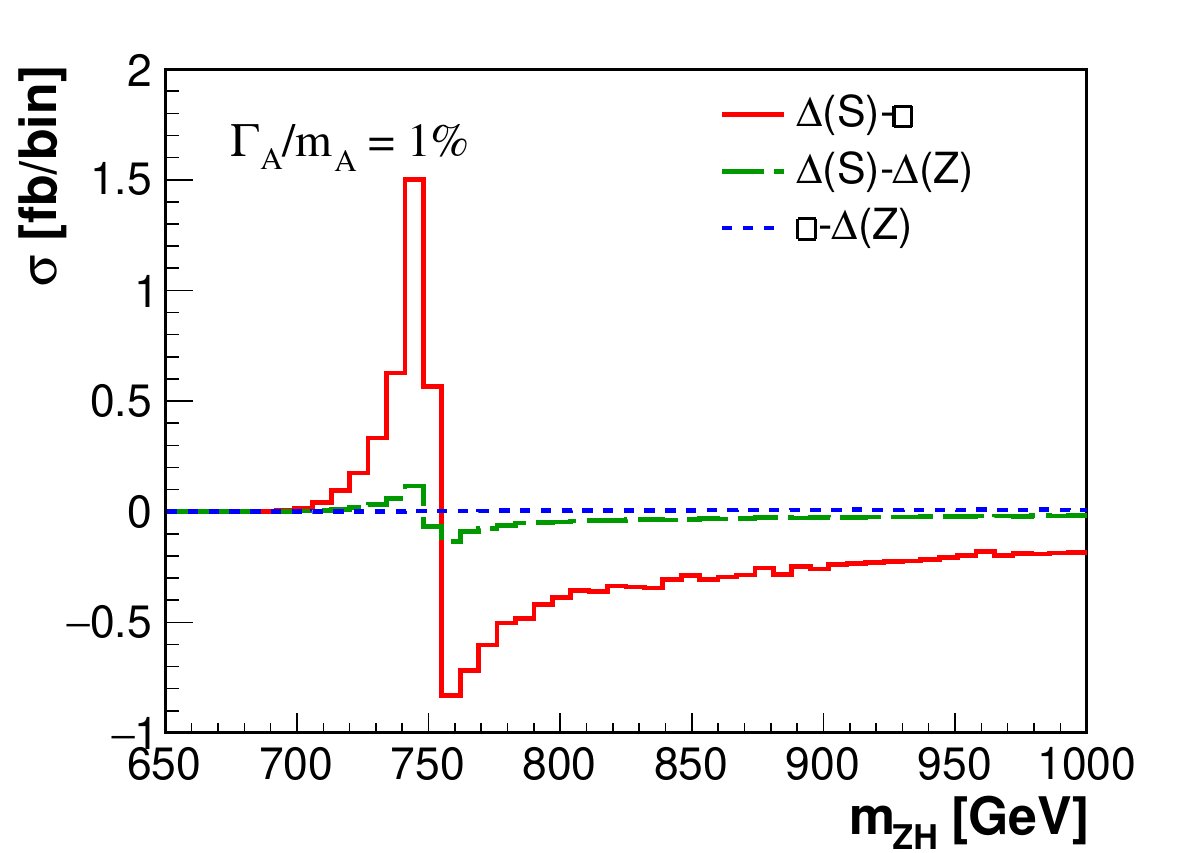}
    \includegraphics[width=0.48\textwidth]{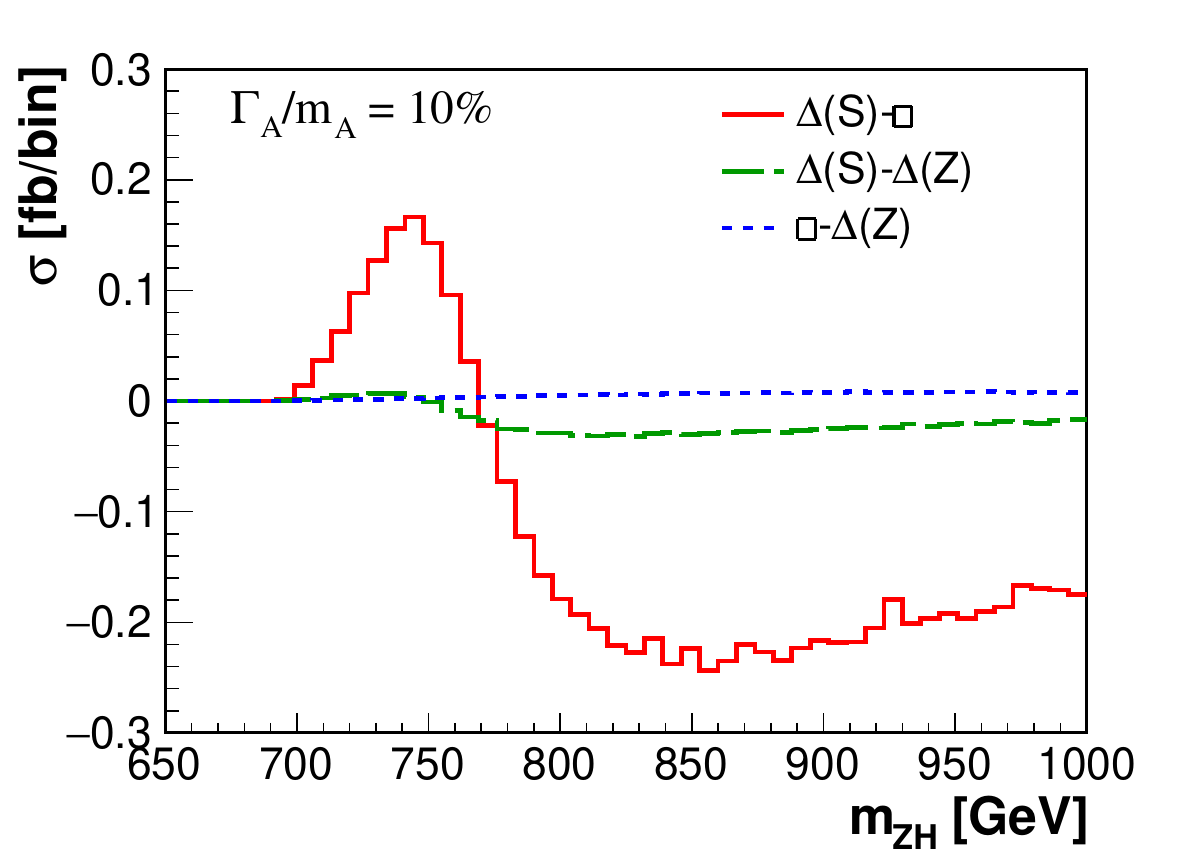}
    \caption{The distributions of $m_{ZH}$ for $gg\to ZH$, with $m_A=750$ GeV, $m_H=600$ GeV, $c_{\beta-\alpha}=0.1$, and $t_\beta = 1$ in the type-I 2HDM. The width of $A$ is chosen to be $1\%$ ($10\%$) of the mass in left (right) panels. In the upper panels, the red, green and blue lines correspond to contributions from the individual set of diagrams presented by~\autoref{fig:AZH_FeynDiag}~(a), \autoref{fig:AZH_FeynDiag}~(b) and~\autoref{fig:AZH_FeynDiag}~(c), respectively. In the lower panels, red, green and blue lines represent the contributions from the interference between different set of diagrams as indicated by the legend.}
    \label{fig:parton_mZH}
\end{figure}

Although resonant production $gg\to A/H\to ZH/A$ typically results in high rates, the other gluon fusion contributions present  relevant effects that need to be included for a robust simulation. In particular, the triangle and box diagrams depicted in~\autoref{fig:AZH_FeynDiag}~(a) and \autoref{fig:AZH_FeynDiag}~(b) display sizable interference effects, which significantly depend on the scalar and pseudo-scalar particle widths. To illustrate it, we present in~\autoref{fig:parton_mZH} the parton level distributions $m_{ZH}$ obtained for a benchmark point with $m_H = 600$~GeV, $m_A = 750$~GeV, $c_{\beta-\alpha}=0.1$, and $t_\beta = 1$. The widths for $A$ and $H$ are chosen to be $\Gamma_{H,A}/m_{H,A}=1\%$ (left panel) and 10\% (right panel). We present the separate components of the signal process in the top panel, including triangle diagrams mediated by a $A$-mediated resonance only (red), box diagrams only (green), $Z$-mediated diagrams only (blue). The respective interferences contributions are shown in the bottom panels of the same figure. The resonant production $gg\to A\to ZH$ results in leading effects, followed by its interference with the box contributions. In general, there is also interference between the signal with the $t\bar{t} Z$ background. However, as checked numerically, such interference generates only subleading rates for the allowed 2HDM parameter space.

\section{Analysis}
\label{sec:ana}
In this section, we derive the sensitivity to the Higgstrahlung signals $pp\to Z(\ell\ell)H(t\bar t)$ and $pp\to Z(\ell\ell)A(t\bar t)$ at the $\sqrt{s}=14$~TeV HL-LHC. The analysis focus on the semi-leptonic top pair final state. The leading background for this process arises from $t\bar t Z$ production. In our study, the Monte Carlo event generation is performed with {\tt MadGraph5aMC@NLO\_v3.1.1}~\cite{Alwall:2014hca}, where the 2HDM model file is prepared with {\tt FeynRules}~\cite{Alloul:2013bka} and {\tt NLOCT}\cite{Degrande:2014vpa}.\footnote{One can find the UFO model files for 2HDM at~\url{https://github.com/ycwu1030/2HDM_FR}.} This model allows the computation of tree-level and one-loop amplitudes. The hadronization and underlying event effects are simulated with {\tt Pythia\_v8.306}~\cite{Sjostrand:2014zea}. Detector effects are accounted for by {\tt Delphes\_v3.5.0}~\cite{deFavereau:2013fsa} package.

The simulation is performed on a grid spanned in four parameters $(m_A,m_H,\Gamma_A,\Gamma_H)$, {setting $t_\beta=1$ and $c_{\beta-\alpha}=0.1$. The proper values of $t_\beta$ and $c_{\beta-\alpha}$  will be adjusted later, satisfying the scan on the parameters of \autoref{eq:scan}}.
For $m_A$ and $m_H$, we scan the light scalar mode within $[400,750]$~GeV with 50~GeV step size, while the more massive one will be at least 100~GeV heavier to allow for the decay $A/H\to ZH/A$ and within $[500,850]$~GeV with 50~GeV step size. For the widths, we follow the same grid adopted by CMS in Ref.~\cite{Sirunyan:2019wph} for their $gg\to H/A\to t\bar{t}$ analysis, such that $\Gamma_{A/H}/m_{A/H}=0.5\%,1.0\%,2.5\%,5.0\%,10.0\%,25.0\%$. For any other values of masses and widths obtained in our parameter scan, interpolation will be used to extract the corresponding results.
To easily cover all other possible choices of  parameters, the  generation of resonant and interference terms for the signal is performed separately for contributions with distinct dependence on $t_\beta$ and $c_{\beta-\alpha}$. Hence, for any other case, we can scale the result according to the corresponding dependence on $t_\beta$ and $c_{\beta-\alpha}$.
The dependence of each diagram on $t_\beta$ and $c_{\beta-\alpha}$ is listed in~\autoref{tab:tb_cba_dependence}.\footnote{The dependence of the $A$ and $H$ widths with $t_\beta$ and $c_{\beta-\alpha}$ is not included here. The effects from the change of the width have already been taken into account when we perform the simulation on the grid and interpolate the results.}
The procedure described in this paragraph will allow for a more effective scan of the model parameters. In particular, this will be relevant for the broad parameter space scan of the type-I 2HDM performed in \autoref{sec:EWPT}.

\begin{table}
    \centering
    \resizebox{\textwidth}{!}{
    \begin{tabular}{cccccccc}
    \hline\hline
    Contribution & \autoref{fig:AZH_FeynDiag}(a) & \autoref{fig:AZH_FeynDiag}(b) & \autoref{fig:AZH_FeynDiag}(c) & \autoref{fig:AZH_FeynDiag}(d) & \autoref{fig:AZH_FeynDiag}(e) & \autoref{fig:AZH_FeynDiag}(f) & \autoref{fig:AZH_FeynDiag}(g) \\
    \hline
    $\kappa$'s for $ZH$ & $\kappa_f^A\kappa'_V\kappa_f^H$ & $\kappa_f^H\kappa_f^H$ & $\kappa_V\kappa_f^H$ & $\kappa_V\kappa_f^H$ & $\kappa_V\kappa_f^H$ & $\kappa_f^A\kappa'_V\kappa_f^H$ & $\kappa_f^H\kappa_f^H$ \\
    \hline
    $\kappa$'s for $ZA$ & $\kappa_f^H\kappa'_V\kappa_f^A$ & $\kappa_f^A\kappa_f^A$ & {$\kappa_V\kappa_f^A$} & - & {$\kappa_V\kappa_f^A$} & $\kappa_f^H\kappa'_V\kappa_f^A$ & $\kappa_f^A\kappa_f^A$\\
    \hline\hline
    \end{tabular}
    }
    \caption{Couplings involved in the diagrams of~\autoref{fig:AZH_FeynDiag}, where $\kappa_f^A=1/t_\beta$, $\kappa_f^H=s_\alpha/s_\beta$, $\kappa_V=c_{\beta-\alpha}$, and $\kappa'_V=s_{\beta-\alpha}$.}
    \label{tab:tb_cba_dependence}
\end{table}

In our event analysis, we require three isolated leptons with  $p_{T\ell}>20$~GeV and $|\eta_\ell|<2.5$.  We require one charged lepton pair of the same flavor and opposite sign, whose invariant mass well reconstructs the $Z$-boson,  $|m_{\ell\ell}-m_Z|<15$~GeV. For the hadronic part of the event, we demand at least four jets, where two are $b$-tagged. Jets are defined with the anti-$k_T$ jet algorithm with radius $R=0.5$ and $p_{Tj}>30$~GeV using {\tt FastJet}~\cite{Cacciari:2011ma}. As the signal displays a semi-leptonic top pair final state, it results in a considerable amount of missing energy stemming from the missing neutrino. Hence, it is required sizable transverse missing energy $\slashed{E}_T>30$~GeV.

Since our signal can display a resonant behavior on the invariant mass $m_{tt}$, as well as $m_{Ztt}$, our analysis explores the profile of the two-dimensional distribution $(m_{tt},m_{Ztt})$ to achieve a superior signal identification. Thus, we reconstruct the missing neutrino from the leptonic top decay. In our signal events, the measured $\slashed{E}_T$ comes mainly from the transverse momentum of the neutrino. Using the third lepton, we calculate the longitudinal component of the neutrino by using the $W$-boson mass constraint, which in most cases will provide two solutions for the neutrino momentum. The reconstruction of the top-quarks is performed by iterating over all possible partitions of light and $b$-tagged jets, forming the leptonic $(\ell\nu b)$ and  hadronic  $(jjb)$ top-quarks. The two possible neutrino solutions are individually accounted for, which constitute different partitions. We chose the combination that minimizes
\begin{align}
        (m_{\ell\nu b}-m_t)^2+(m_{jj b}-m_t)^2,
\end{align}
where $m_t$ is the on-shell top-quark mass. With the reconstructed top-quarks, we can examine the invariant masses $m_{tt}$ and $m_{Ztt}$, which capture the resonant profile from  $H$ and $A$ decays, arising from the loop-induced gluon fusion $gg\to A/H\to ZH/A$ and bottom quark-initiated $b\bar{b}\to A/H\to ZH/A$ contributions.

\begin{figure}[!tb]
    \centering
    \includegraphics[width=0.32\textwidth]{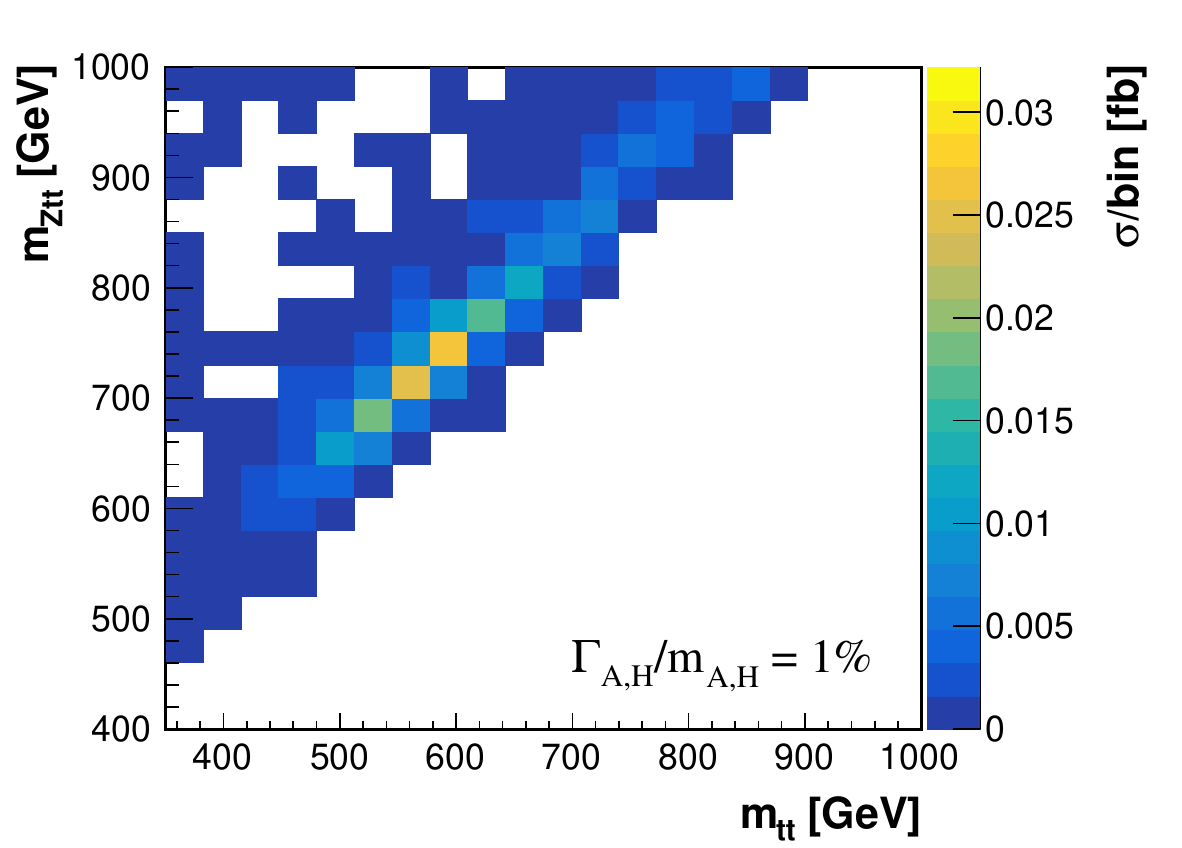}
    \includegraphics[width=0.32\textwidth]{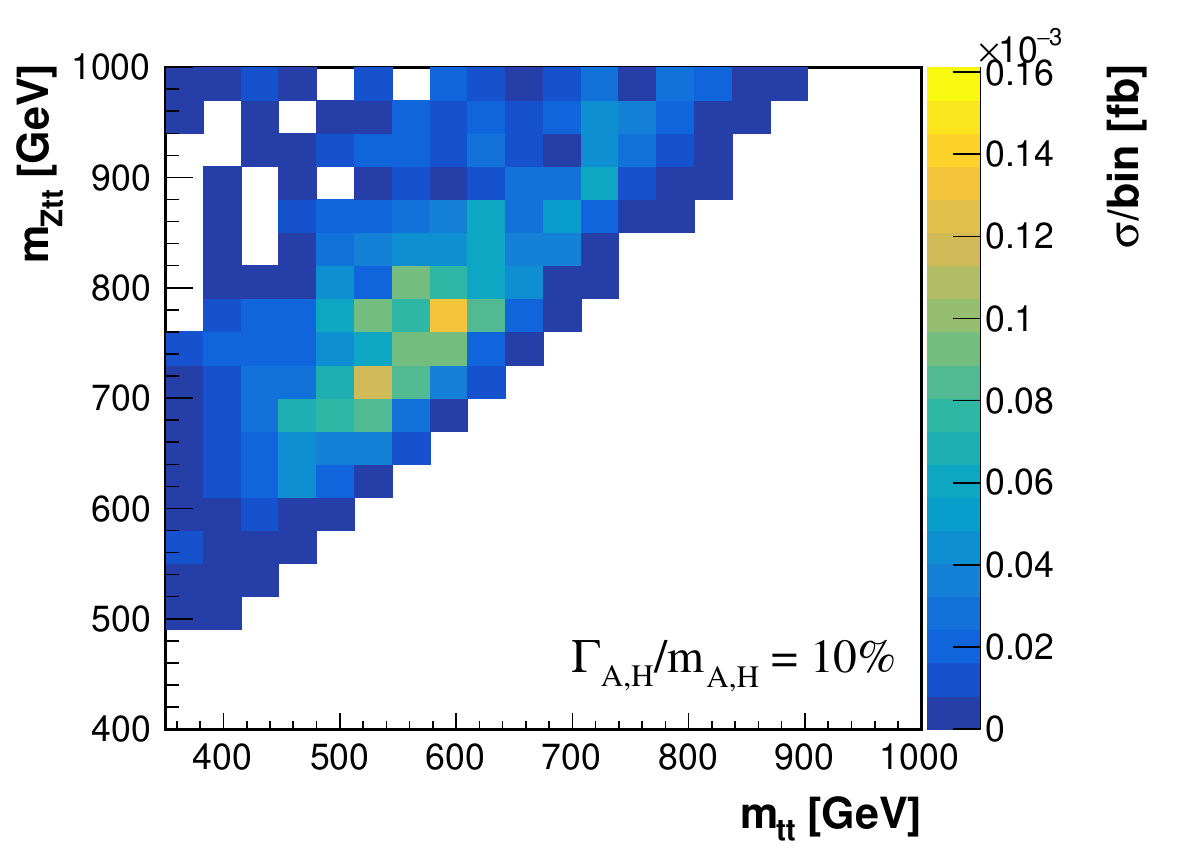}
    \includegraphics[width=0.32\textwidth]{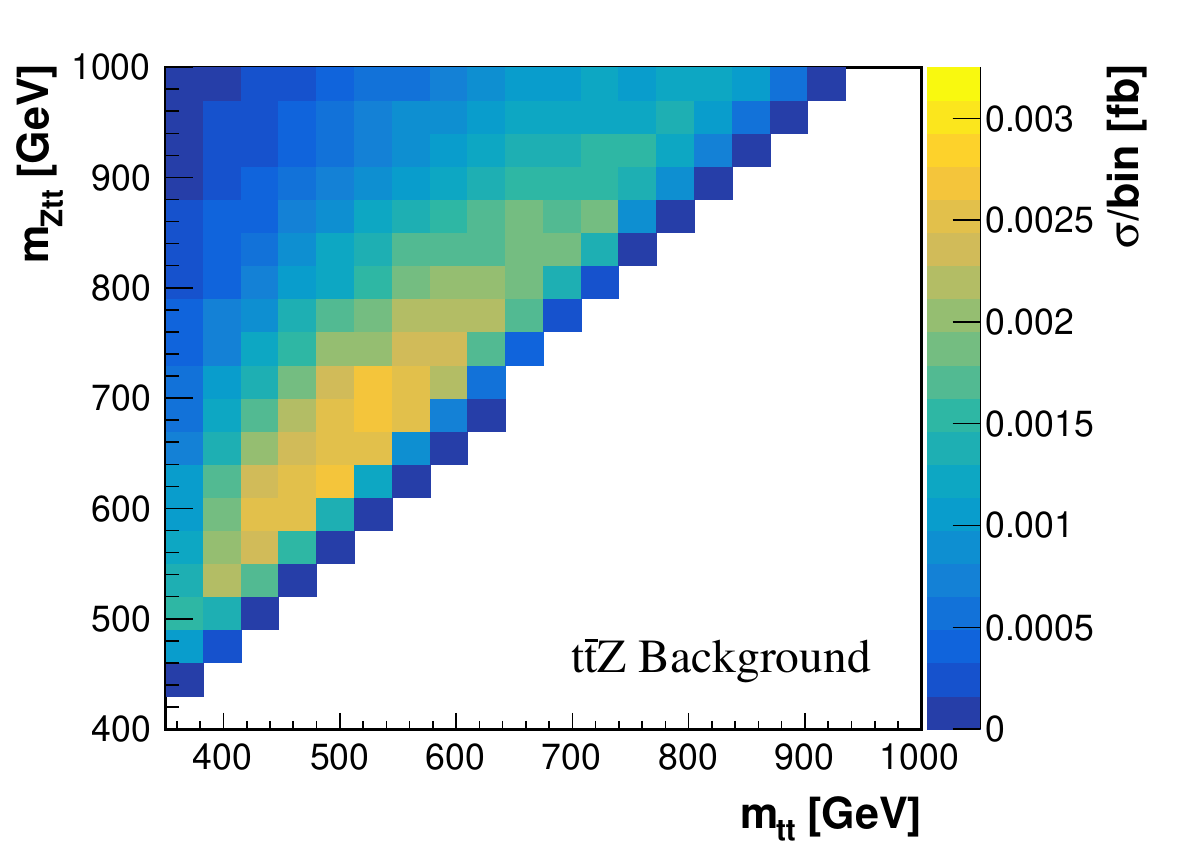}
    \caption{The 2D distribution in $m_{tt}$-$m_{Ztt}$ plane for $m_A=750$ GeV and $m_H=600$ GeV with $\Gamma_{A,H}/m_{A,H}=1\%$ (left panel) and $\Gamma_{A,H}/m_{A,H}=10\%$ (middle panel). The $t\bar t Z$ background is also presented (right panel). The cross-section around the white pixels are either not kinematically allowed (right region) or suppressed (top left region), and we could not find any events in our scan.}
    \label{fig:2D_dist}
\end{figure}

\begin{figure}[!tb]
    \centering
    \includegraphics[width=\textwidth]{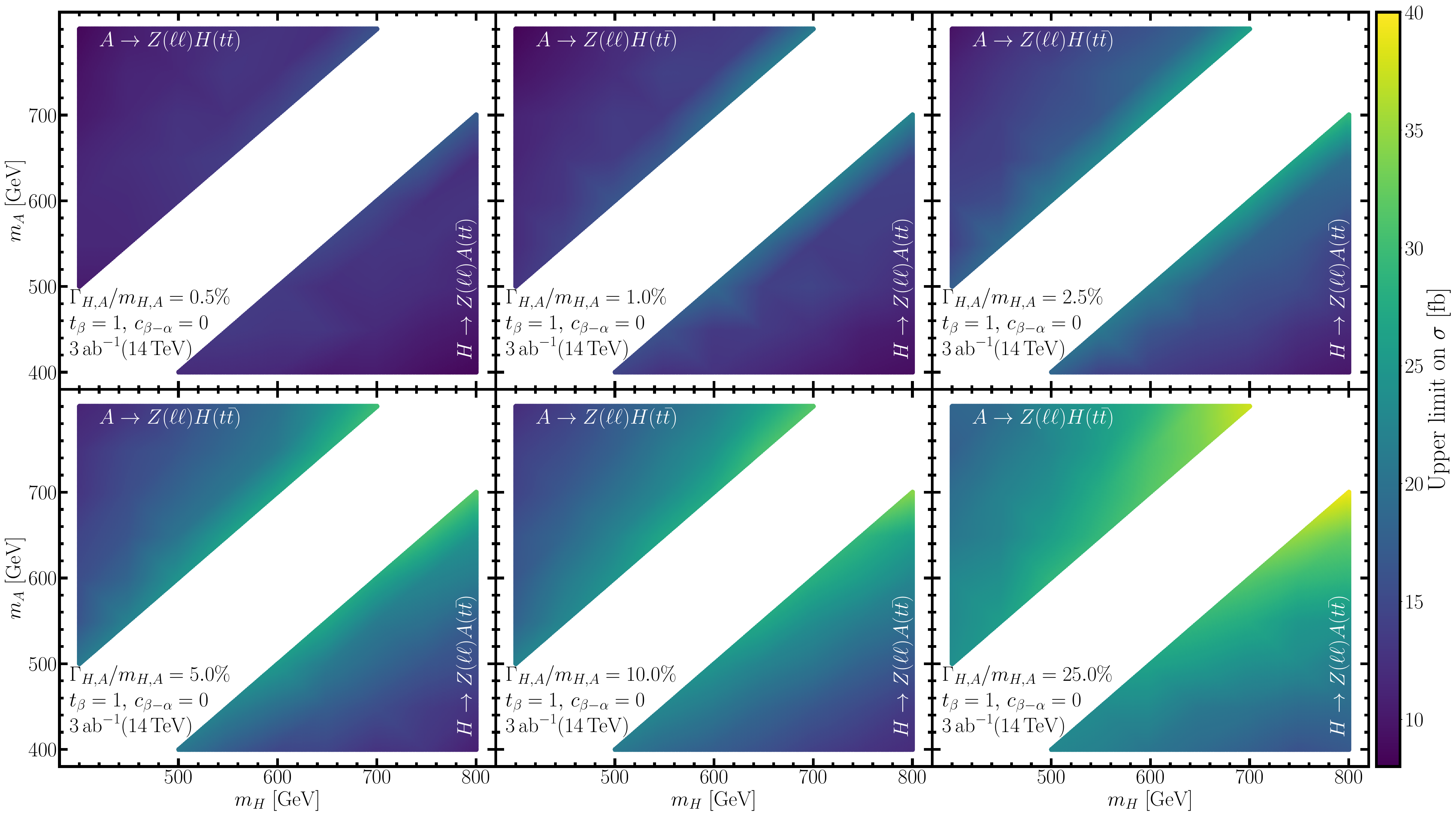}
    \caption{95\% CL upper limit on the cross-section $\sigma(pp\to ZH/A)\times{\rm BR}(H/A\to t\bar t)$ in the $m_A$-$m_H$ plane, with different choice of  widths for the $H$ and $A$ bosons. We consider the $\sqrt{s}=14$~TeV HL-LHC with 3~ab$^{-1}$ of integrated luminosity.}
    \label{fig:95CL_mu}
\end{figure}

To estimate the high luminosity LHC sensitivity to Higgstrahlung signals $pp\to Z(\ell\ell)H(t\bar t)$ and $pp\to Z(\ell\ell)A(t\bar t)$, we performed a binned log-likelihood analysis based on the two-dimensional distribution $(m_{tt},m_{Ztt})$. For illustration, the 2D distribution in $m_{tt}$-$m_{Ztt}$ plane for $m_{A}=750$~GeV and $m_H=600$~GeV is shown in~\autoref{fig:2D_dist} with $\Gamma_{A,H}/m_{A,H}=1\%$ and $\Gamma_{A,H}/m_{A,H}=10\%$. The correspondent $t\bar t Z$ background distribution is also shown. In~\autoref{fig:95CL_mu}, we present the resulting upper limit on the signal cross-section for $Z(\ell\ell)H(t\bar t)$ and $Z(\ell\ell)A(t\bar t)$ production at 95\% confidence level (CL). The results are shown for several hypotheses of scalar and pseudoscalar widths.\footnote{More results can be found at \url{https://github.com/ycwu1030/AZH_Worker}, where the analysis is saved.} We assume the HL-LHC with integrated luminosity $\mathcal{L}=3$~ab$^{-1}$. The binned likelihood analysis is sensitive to the width dependence, leading to weaker results in the augmented $\Gamma_{H,A}/m_{H,A}$ regime. This is mostly a result of the enhanced negative contribution from the triangle-box interference for larger widths. The large interference effects and broader resonance from large width lead to suppressed signal events, which can be observed in~\autoref{fig:parton_mZH} and~\autoref{fig:2D_dist}. While we present the results in Fig.~\ref{fig:95CL_mu} following a model-independent approach similar to the CMS study in Ref.~\cite{Sirunyan:2019wph}, in Sec.~\ref{sec:results} we perform a uniform parameter space scan considering the 2HDM in the type-I scenario.

\section{Electroweak Phase Transition in the 2HDM}
\label{sec:EWPT}

In this section, we study the complementarity between collider and gravitational wave experiments to probe the phase transition profile in the 2HDM. We first define the one-loop effective potential at finite temperature in~\autoref{sec:eff-pot}, then discuss important ingredients for calculating the EWPT and gravitational wave sensitivity at space-based experiments in~\autoref{sec:history}. Finally, we analyze in~\autoref{sec:results} the relevance of the Higgstrahlung channel with top pair final states $H/A\to t\bar t$ with respect to other relevant classes of searches at the HL-LHC and also contrast the collider with the gravitational wave sensitivity.

\subsection{Finite Temperature Effective Potential}
\label{sec:eff-pot}

In order to study the thermal history of the 2HDM, one needs to study the loop-corrected effective potential at finite temperature. It is defined by the addition of the tree-level potential $V_0$, the zero-temperature one-loop corrections from the Coleman-Weinberg potential $V_{CW}$ with the respective counter terms $V_{CT}$, and the one-loop thermal corrections $V_T$. Thus, the final potential can be written as
\begin{align}
    V_\text{eff}=V_0+V_{CW}+V_{CT}+V_T.
\end{align}

The zero-temperature one-loop correction in the $\overline{\text{MS}}$ have the form~\cite{PhysRevD.7.1888}
\begin{align}
    V_{CW} &= \sum_i \frac{n_i}{64\pi^2}m_i^4(\Phi_1,\Phi_2)\left[\log \left(\frac{m_i^2(\Phi_1,\Phi_2)}{\mu^2}\right)-c_i\right]\,,
    \label{eq:CW}
\end{align}
where $m_i(\Phi_1,\Phi_2)$ is the field-dependent mass of the particle of type $i$, with $n_i$ being the correspondent number of degrees of freedom. We define $n_i>0$ for bosons and $n_i<0$ for fermions. The renormalization scale constant is set to the electroweak VEV at zero-temperature, $\mu=v\approx 246$~GeV. The $c_i$ are the renormalization constants in the $\overline{\text{MS}}$ scheme, with $c_i=5/6$ for gauge bosons and $c_i=3/2$ for scalars and fermions. The Coleman-Weinberg potential was evaluated in the Landau gauge, allowing for the omission of ghost contributions.

In principle, the tree-level scalar masses and mixing angles are shifted by the addition of the Coleman-Weinberg potential. To make our parameter space scan more efficient, we follow a renormalization prescription that requires these parameters to match their tree-level values~\cite{Camargo-Molina:2016moz,Basler:2016obg}. We achieve this by adding the counter-terms
\begin{align}
	V_{CT} =& \delta m_{11}^2\Phi_1^\dagger\Phi_1 + \delta m_{22}^2\Phi_2^\dagger\Phi_2 - \delta m_{12}^2(\Phi_1^\dagger\Phi_2 + h.c.) + \frac{\delta \lambda_1}{2}(\Phi_1^\dagger\Phi_1)^2 + \frac{\delta\lambda_2}{2}(\Phi_2^\dagger\Phi_2)^2 \nonumber \\
            & + \delta\lambda_3(\Phi_1^\dagger\Phi_1)(\Phi_2^\dagger\Phi_2) + \delta\lambda_4(\Phi_1^\dagger\Phi_2)(\Phi_2^\dagger\Phi_1) + \frac{\delta\lambda_5}{2}\left((\Phi_1^\dagger\Phi_2)^2 + h.c.\right)\,,
\label{eq:CT}
\end{align}
imposing the following renormalization conditions at zero-temperature
\begin{align}
\partial_{\phi_i}(V_{CW}+V_{CT})|_{\omega=\omega_\text{tree}}=0 \,,
\label{eq:ren1}\\
\partial_{\phi_i}\partial_{\phi_j}(V_{CW}+V_{CT})|_{\omega=\omega_\text{tree}}=0\,,
\label{eq:ren2}
\end{align}
where $\omega_\text{tree}$ generically stands for the minimum of the tree-level potential for the $\phi_i$ fields.
\autoref{eq:ren1} requires that the zero-temperature minimum matches the tree-level value. Similarly, \autoref{eq:ren2} demands that $T=0$ masses and mixing angles do not change with respect to the tree-level numbers.

The one-loop thermal corrections read~\cite{Arnold:1992rz}
\begin{align}
    V_T&=\frac{T^4}{2\pi^2} \left[
    \sum_f n_f J_+\left(\frac{m_f^2}{T^2}\right)+
    \sum_{\mathcal{V}_T} n_{\mathcal{V}_T} J_-\left(\frac{m_{\mathcal{V}_T}^2}{T^2}\right)
    +\sum_{\mathcal{V}_L} n_{\mathcal{V}_L}  J_-\left(\frac{m_{\mathcal{V}_L}^2}{T^2}\right) \right] \nonumber\\
    &-\frac{T^4}{2\pi^2}\sum_{\mathcal{V}_L} \frac{\pi}{6} \left(\frac{\overline{m}^3_{\mathcal{V}_L}}{T^3}-\frac{m_{\mathcal{V}_L}^3}{T^3}\right) \,,
    \label{eq:VT}
\end{align}
where $f$, $\mathcal{V}_T$, and $\mathcal{V}_L$ indicate respectively the sum over fermions, transverse gauge bosons ($W_T,Z_T$), and longitudinal modes of gauge bosons and scalars ($W_L,Z_L,\gamma_L,\Phi^0,\Phi^{\pm}$). The second line of~\autoref{eq:VT} corresponds to the daisy contributions, following the Arnold-Espinosa scheme~\cite{Arnold:1992rz,Basler:2016obg}.
Finally, the thermal functions $J_+$ and $J_-$ are given by
\begin{align}
    J_{\pm}(x)=\mp\int_0^\infty dy~ y^2 \log\left(1\pm e^{-\sqrt{y^2+x^2}} \right)\,.
\end{align}

\subsection{Thermal History and Gravitational Waves}
\label{sec:history}

The rich structure of the 2HDM potential grants distinct phase transition processes. For successful baryogenesis, it is critical to prevent the baryon number generated during the phase transition from being significantly washed out. This imposes the electroweak phase transition to be strong first-order~\cite{Moore:1998swa,Quiros:1999jp}
\begin{equation}
 \xi_c\equiv\frac{v_c}{T_c}\gtrsim 1,
\end{equation}
where $T_c$ is the critical temperature and $v_c$ is the critical Higgs VEV obtained when the would-be true vacuum and false vacuum are degenerate.

The occurrence of phase transition depends on the false to true vacuum tunneling rate, which is given by~\cite{Linde:1980tt,Coleman:1977py}
\begin{equation}
\Gamma (T)\approx T^4\left(\frac{S_3}{2\pi T}\right)^{3/2}e^{-\frac{S_3}{T}},
\end{equation}
where $S_3$ denotes the three-dimensional Euclidean action that can be expressed as
\begin{equation}
    S_3=4\pi\int_{0}^{\infty}{dr r^2\left [ \frac{1}{2}
    \left (\frac {d\phi(r)}{dr} \right )^2+V(\phi,T) \right ]}\,.
\end{equation}
To obtain the nucleation rate, we need to calculate the scalar field $\phi$ bubble profile by solving the bounce equation
\begin{equation}
    \frac {d^2\phi}{dr^2}+\frac{2}{r} \frac{d\phi}{dr}=\frac{dV(\phi,T)}{d\phi},
\end{equation}
with boundary conditions
\begin{equation}
    \lim_{r\rightarrow \infty}\phi(r)=0
    \quad \text{and} \quad \lim_{r\rightarrow 0}\frac {d\phi(r)}{dr}=0.
\end{equation}
We solve the differential equation and calculate $S_3$ with {\tt CosmoTransitions v2.0.5}~\cite{Wainwright:2011kj}. The nucleation onsets at temperature $T_n$, where one bubble nucleates per horizon volume. This condition can be approximated for EWPT as $S_3(T_n)/T_n\approx 140$~\cite{Moreno:1998bq}.

We can now define the quantities $\beta$ and $\alpha$ that describe the dynamical properties of the phase transition, which can be used to model the strength of stochastic gravitational wave signals~\cite{Grojean:2006bp,Caprini:2015zlo}. The inverse time duration of phase transition $\beta/H_n$ is defined as
\begin{align}
	\frac{\beta}{H_n} \equiv T_n\frac{d}{dT}\left.\left(\frac{S_3}{T}\right)\right|_{T=T_n},
\end{align}
where $H_n$ is the Hubble constant at $T_n$. The other important parameter $\alpha$ characterizes the ratio between the latent heat released during the phase transition ($\epsilon$) with respect to the radiation energy density ($\rho_{rad}$), $\alpha\equiv\epsilon/\rho_{rad}$. These parameters are given by
\begin{align}
 \epsilon = \Delta\left(- V_{\rm eff} + T\frac{\partial V_{\rm eff}}{\partial T}\right)_{T=T_n} \quad \text{and} \quad \rho_{\rm rad} = \frac{\pi^2}{30}g_\star T_n^4\,,
\end{align}
with $g_\star$ being the number of relativistic degrees of freedom in the thermal plasma. Here, $\Delta$ denotes the difference between true and false vacua.

We are able to estimate the sensitivity of gravitational wave experiments, adopting the signal-to-noise ratio (SNR)  measure~\cite{Caprini:2015zlo}
\begin{align}
\mathrm{SNR}=\sqrt{\mathcal{T} \int_{f_{\min }}^{f_{\max }} d f\left[\frac{h^{2} \Omega_{\mathrm{GW}}(f)}{h^{2} \Omega_{\mathrm{Sens}}(f)}\right]^{2}}\,,
\end{align}
where $\mathcal{T}$ is the duration of the mission and $\Omega_{\rm Sens}$ is the sensitivity profile of the particular GW experiment~\cite{Audley:2017drz}. In our analysis, we consider the space-based LISA experiment, assuming $\mathcal{T}=5$ years and  ${\rm SNR}>10$ to characterize signal detection~\cite{Caprini:2015zlo}.

\subsection{Probing EWPT in the 2HDM}
\label{sec:results}

In this section, we examine the sensitivity of the resonant top pair searches to the parameter space regime related to SFOEWPT in the 2HDM. Instead of performing a model-independent analysis as in \autoref{fig:95CL_mu}, we implement a uniform parameter space scan of the 2HDM for the type-I scenario. The  uniformly random scan is performed over the region
\begin{align}
    \tan\beta &\in (0.8,25)\,,           &m_{12}^2&\in(10^{-3},10^5)\,{\rm GeV}^2\,, &m_H&\in(150,1500)\rm\,GeV \,,    \nonumber \\
    \cos(\beta-\alpha)&\in(-0.3,0.3)\,,  & m_A&\in(150,1500)\,{\rm GeV}\,,           &m_{H^\pm}&\in(150,1500)\,{\rm GeV}.
     \label{eq:param_scan}
\end{align}
The theoretical and experimental limits to the model are implemented with {\tt ScannerS v2.0.0} \cite{Coimbra:2013qq,Muhlleitner:2020wwk}. With this package, we require constraints from perturbative unitarity~\cite{Lee:1977eg,Kanemura:1993hm,Ginzburg:2005dt}, boundness from below~\cite{Ivanov:2018jmz}, vacuum stability~\cite{Hollik:2018wrr,Ferreira:2019iqb}, flavor constraints, and electroweak precision. {\tt HiggsBounds v5.3.2} and {\tt HiggsSignals v2.2.3} are used to account for the heavy Higgs searches and the 125~GeV Higgs boson measurements~\cite{Bechtle:2020pkv,Bechtle:2020uwn}.

\begin{figure}
    \centering
    \includegraphics[width=\textwidth]{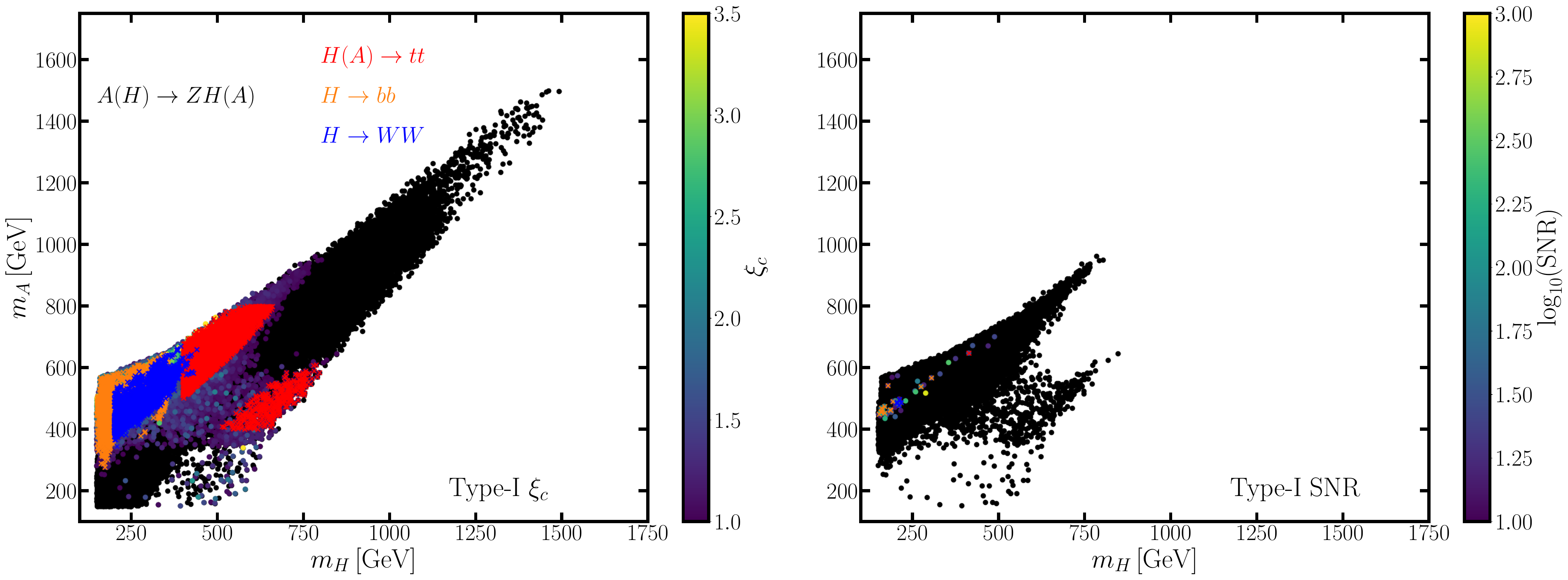}
    \caption{The $pp\to ZH/A$ constraints on $m_H$-$m_A$ plane. The heat map tracks $\xi_c$ (left panel) and SNR (right  panel). For the left panel, the black points represent the parameter space regime with first-order phase transition with $\xi_c>0$. For the right  panel, the black points indicate the regime with $\xi_c>1$. The parameter space scan is performed with {\tt ScannerS}~\cite{Muhlleitner:2020wwk}, where the constraints from perturbative unitarity, boundness from below, vacuum stability, electroweak precision, and flavor constraints are imposed. {\tt HiggsBounds} and {\tt HiggsSignals} are also included to incorporate the searches for additional scalars as well as the 125 GeV Higgs boson measurement~\cite{Bechtle:2020pkv,Bechtle:2020uwn}. The colored crosses are  points with  $\xi_c>1$ that can be probed by the HL-LHC through $pp\to ZH/A$ ($Z\to\ell\ell$) channel with $H/A\to t\bar t$ (red), $H\to b\bar b$ (orange), and $H\to WW$ (blue).}
      \label{fig:bound}
\end{figure}

The current experimental Higgstrahlung studies $pp\to ZH/A$ only account for the scalar decays  $H/A\to b\bar b $ and  $H\to WW$, associated with $Z\to \ell \ell$~\cite{Sirunyan:2019wrn,Aad:2020ncx}. In~\autoref{fig:bound}, we present the projection for these analyses to the HL-LHC luminosity $\mathcal{L}=3$~ab$^{-1}$. The projected upper bounds on the cross-section for corresponding processes are compared with the cross-section times the branching fraction of each parameter point obtained from {\tt ScannerS}~\cite{Muhlleitner:2020wwk}. Both the bottom and $W$-boson final states will provide relevant sensitivity to SFOEWPT. Interestingly, the sensitivity of these channels is mostly restricted to the $m_{H,A}<350$~GeV regime. Above the top-quark pair threshold, the $H/A\to t\bar t$ final state is typically the dominant scalar decay, favoring phenomenological analyses with a resonant top-quark pair final state. We present in~\autoref{fig:bound} (red points) the results of our $pp\to Z(\ell\ell)H(t\bar t)$ and $pp\to Z(\ell\ell)A(t\bar t)$ analyzes, as described in~\autoref{sec:AZH}. We observe that the Higgstrahlung searches with $t\bar{t}$ final state largely extend the sensitivity to SFOEWPT.

\begin{figure}
    \centering
    \includegraphics[width=0.48\textwidth]{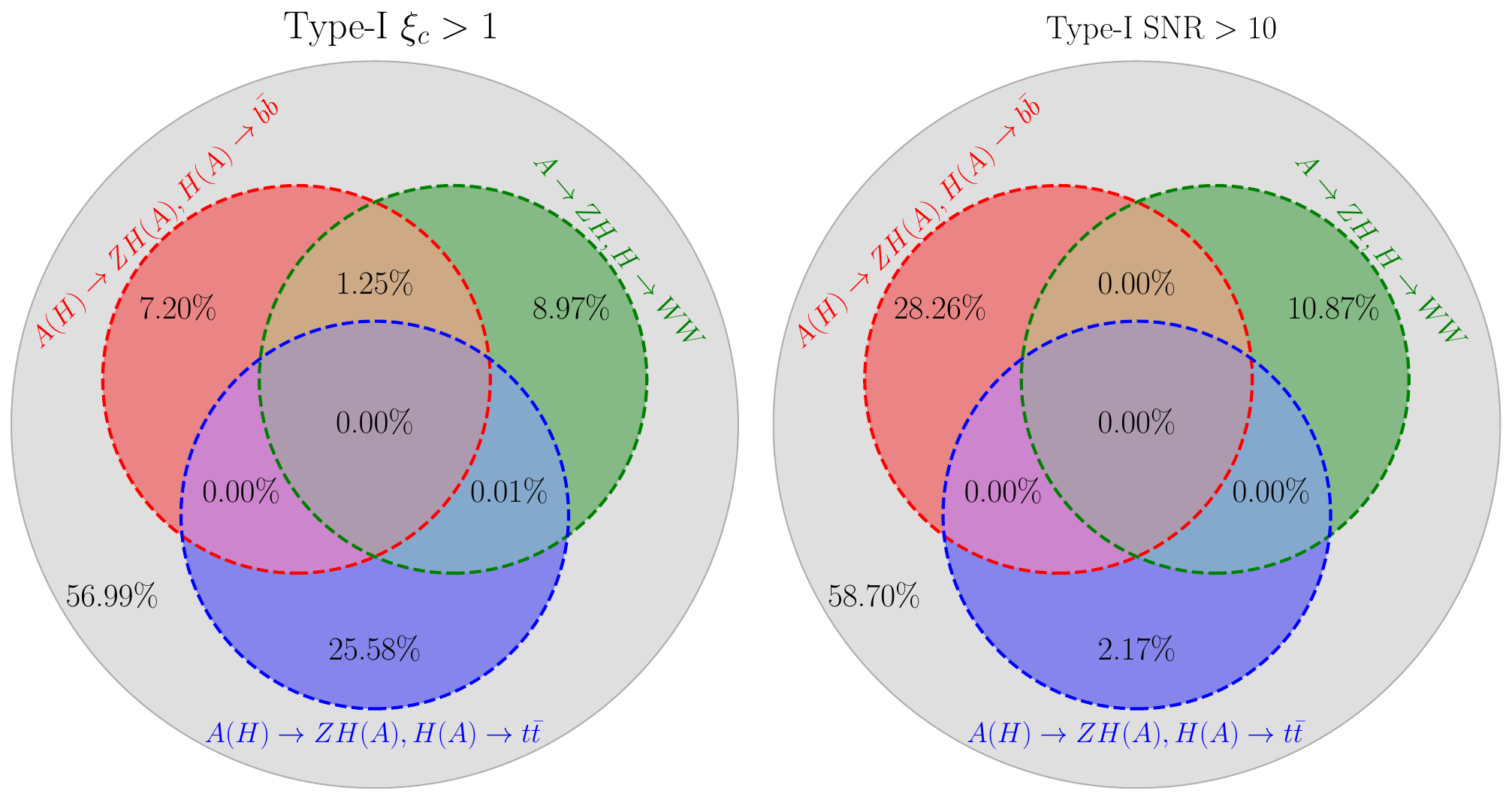}
    \includegraphics[width=0.48\textwidth]{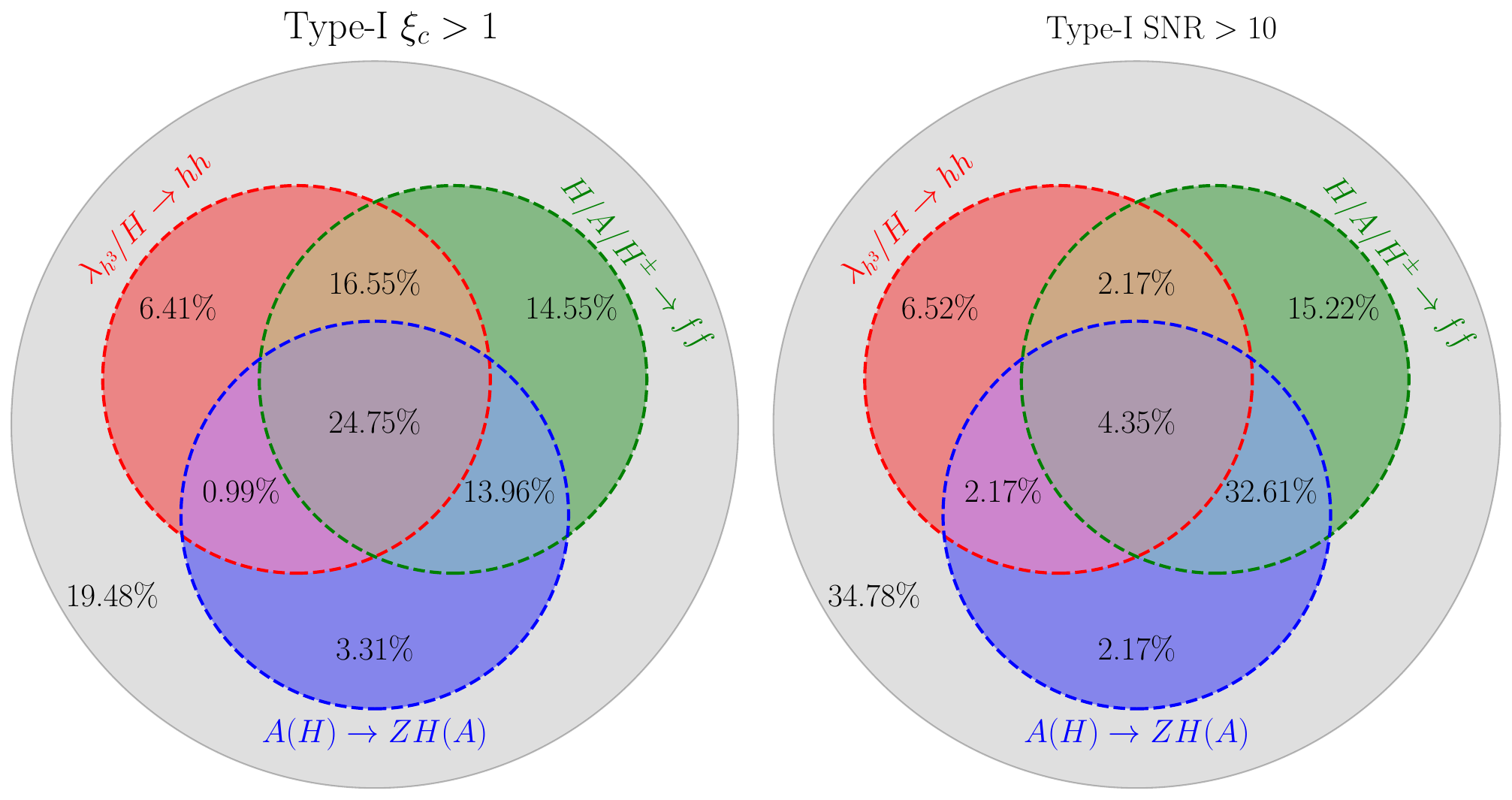}
    \caption{Left: The summary of sensitivities for the Higgstrahlung search channels $pp\to ZH/A$ at the HL-LHC. The number in each region denotes the fraction of currently allowed points, by theoretical and experimental constraints, that will be probed by the Higgstrahlung $b\bar b$, $WW$, and $t\bar t$ final state searches at the HL-LHC.   Right: The summary of the capabilities of corresponding search channels including non-resonant and resonant double Higgs searches $\lambda_{h^3}/H\to hh$, fermionic decay channels $H/A/H^\pm\to ff$, and $A(H)\to ZH(A)$.}
    \label{fig:Ven_bbvsWWvstt}
\end{figure}

In the left two panels of~\autoref{fig:Ven_bbvsWWvstt}, we present a detailed comparison among the $b\bar b$, $WW$, and $t\bar t$ final states for the $pp\to ZH/A$ channel. The number in each region indicates the fraction of points in our uniformly random scan that can be excluded by the corresponding searches. They are the fraction of points currently allowed, under theoretical and experimental constraints, that will be probed at the HL-LHC. The first panel focuses on the SFOEWPT regime with $\xi_c>1$ and the second one represents the points that can display gravitational wave signals at LISA with signal-to-noise ratio $\text{SNR}>10$. We observe from these Venn diagrams that the Higgstrahlung channels with top pair final states $pp\to ZH(t\bar t)$ and $ZA(t\bar t)$ cover a large portion of the allowed parameter space. Therefore, the inclusion of this channel in forthcoming experimental analyses is strongly motivated.

In the right two panels of~\autoref{fig:Ven_bbvsWWvstt}, we show the comparison of the Higgstrahlung sensitivity at the HL-LHC with other relevant channels in the context of EWPT, namely  resonant and non-resonant double Higgs searches $H\to hh$ (or $\lambda_{h^3}$) and $gg\to H/A\to t\bar t$ (or $H^\pm\to tb$) searches. The results from these other studies were obtained from Ref.~\cite{Goncalves:2021egx}. We added further details for the $gg\to H/A\to t\bar{t}$ searches in the~\autoref{sec:ggF}. The HL-LHC will be able to cover $\approx 80\%$ of the available $\xi_c>1$ parameter space in the Type-I 2HDM scenario. While the considered channels present relevant complementarities, the Higgstrahlung mode and gluon fusion scalar production with subsequent decay to top-quark pair (or charged Higgs production with fermionic decay $H^\pm\to tb$) result in leading sensitivities.

\section{Summary}
\label{sec:summary}

The thermal history of electroweak symmetry breaking in the Universe could have profound consequences for particle physics and cosmology. In this article, we study the prospects for the HL-LHC to probe the strong first-order electroweak phase transition regime in the type-I 2HDM. We devote special attention to the Higgstrahlung channel $pp\to ZH/A$ with a resonant top-quark pair final state $H/A\to t\bar t$.

We scrutinize the signal components associated with this channel and find that it is important to go beyond the resonant $gg\to A/H\to ZH/A$ theoretical modeling, including the full signal simulation. The 2HDM can display large interference effects between the resonant and non-resonant $gg\to ZH/A$ production and augmented $b$-quark initiated contributions. Accounting for the theoretical and experimental constraints, we find that the considered Higgstrahlung final state $H/A\to t\bar t$ renders the largest sensitivity to the $\xi_c>1$ regime in the 2HDM, in comparison to the other Higgstrahlung searches already accounted for by ATLAS and CMS, which rely on the $H/A\to b\bar {b}$ and $H\to WW$ final states.

Finally,  we derive the complementarity of the Higgstrahlung searches with other relevant classes of searches at the HL-LHC (namely, resonant and non-resonant double Higgs searches and heavy scalar decays to fermions $A/H/H^\pm\to ff)$ and confront with the gravitational wave sensitivity at  LISA. We obtain that the channel studied here, with top-quark pair final state, will be a promising signature for $\xi_c>1$ at the HL-LHC. Therefore, the inclusion of the studied Higsstrahlung channel in the forthcoming ATLAS and CMS experimental analyses is strongly motivated, as it would both augment the sensitivity to the 2HDM parameter space and substantially boost the sensitivity to SFOEWPT at the LHC.

\appendix
\section{Top Pair Resonant Searches via \texorpdfstring{$gg\to H/A\to t\bar{t}$}{ggtoHAtott}}
\label{sec:ggF}

\begin{figure}
    \centering
    \includegraphics[width=\textwidth]{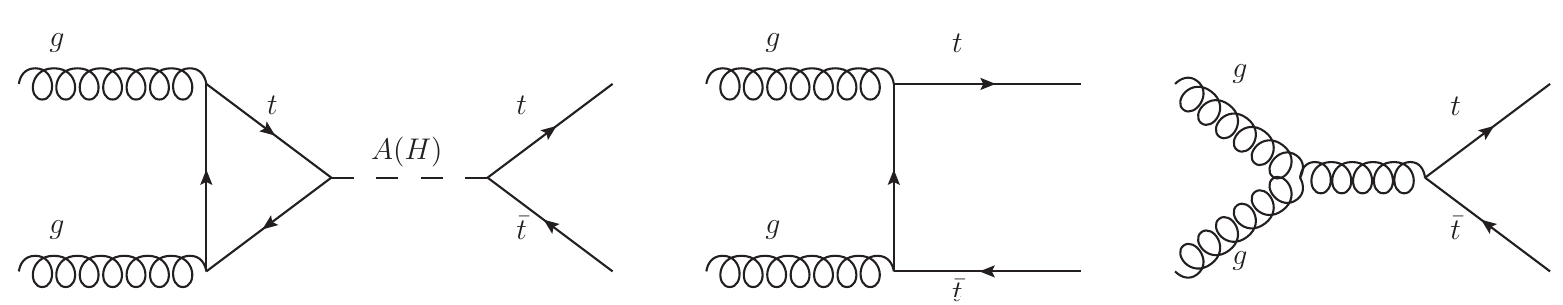}
    \caption{Representative set of Feynman diagrams for the top-quark pair production through scalar resonance $gg\to A/H\to t\bar t$ (left) and directly from gluon fusion $gg\to t\bar t$ (middle and right).}
    \label{fig:gg_tt_feynman}
\end{figure}

In this appendix, we briefly review the top-quark pair resonant searches via gluon fusion $gg\to H/A\to t\bar t$.  The gluon fusion channel displays sizable interference effects between the $s$-channel (pseudo)scalar induced top pair production (left diagram of~\autoref{fig:gg_tt_feynman}) and the QCD top pair production (middle and right diagrams of~\autoref{fig:gg_tt_feynman})~\cite{Gaemers:1984sj,Dicus:1994bm,Bernreuther:1997gs,Barger:2006hm,Frederix:2007gi,Barcelo:2010bm,Figy:2011yu,Barger:2011pu,Moretti:2012mq,Craig:2015jba,Bernreuther:2015fts,Gori:2016zto,Djouadi:2016ack,Hespel:2016qaf,Czakon:2016vfr,Carena:2016npr,Djouadi:2019cbm}. The large interference has significant phenomenological consequences, since the top-quark invariant mass $m_{tt}$ does not display a simple resonance profile around the (pseudo)scalar mass $m_{H/A}$, instead it presents a bump-dip shape.

\begin{figure}
    \centering
    \includegraphics[width=0.48\textwidth]{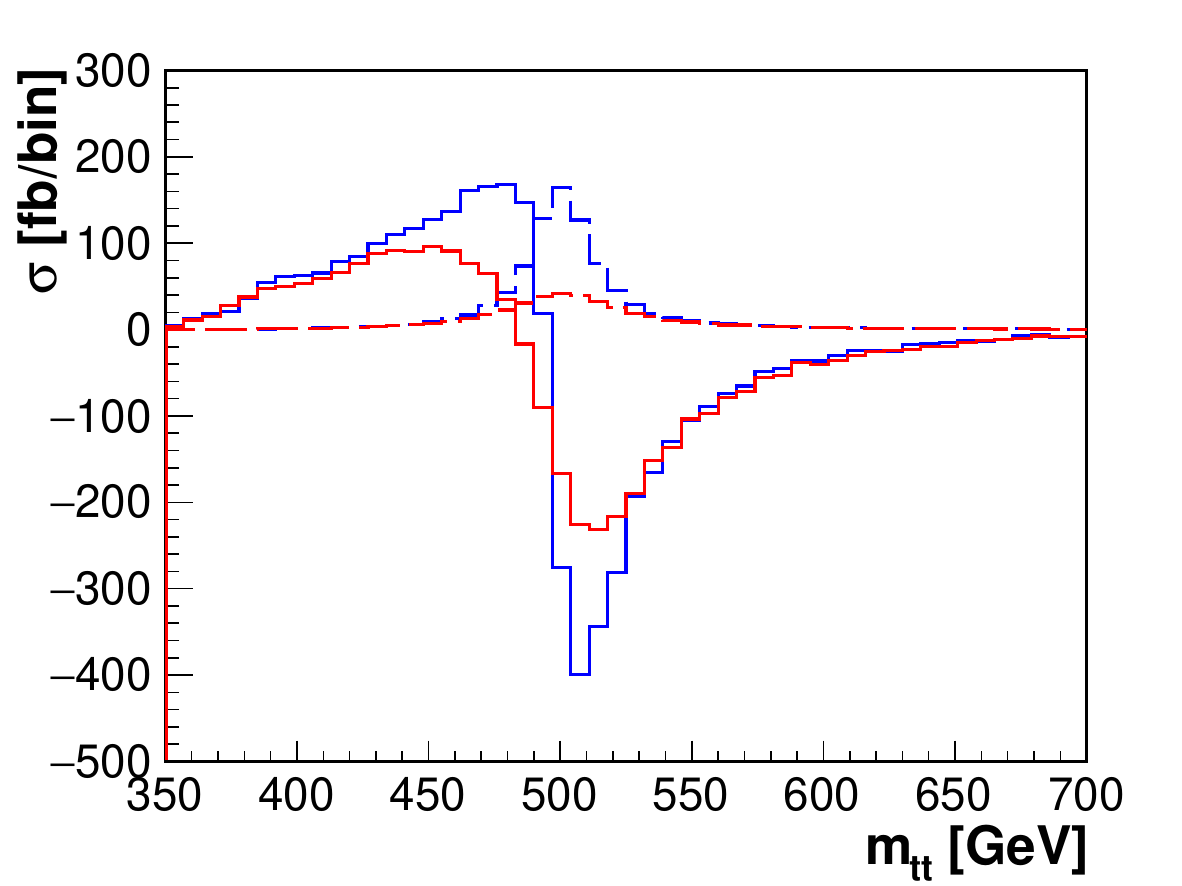}
    \put(-85,125){$gg\to H\to t\bar t$}
    \put(-85,78){\color{blue}\rule{0.025\textwidth}{1pt}}\put(-72,76){\scriptsize 5\% Full}
    \put(-85,65){\color{red}\rule{0.025\textwidth}{1pt}}\put(-72,63){\scriptsize 10\% Full}
    \put(-85,52){\color{blue}\rule{0.010\textwidth}{1pt}}\put(-79,52){\color{blue}\rule{0.010\textwidth}{1pt}}\put(-72,50){\scriptsize 5\% Resonant}
    \put(-85,39){\color{red}\rule{0.010\textwidth}{1pt}}\put(-79,39){\color{red}\rule{0.010\textwidth}{1pt}}\put(-72,37){\scriptsize 10\% Resonant}
    \includegraphics[width=0.48\textwidth]{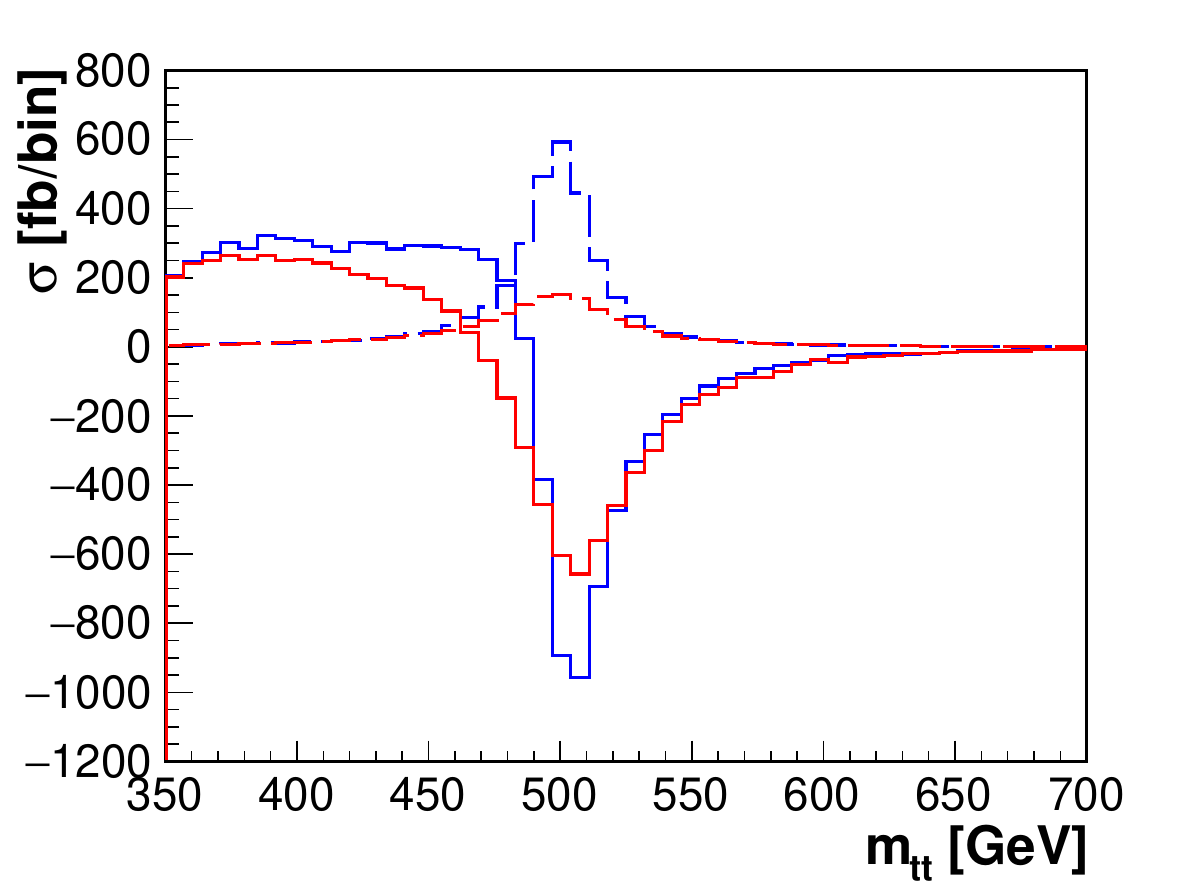}
    \put(-85,125){$gg\to A\to t\bar t$}
    \put(-85,78){\color{blue}\rule{0.025\textwidth}{1pt}}\put(-72,76){\scriptsize 5\% Full}
    \put(-85,65){\color{red}\rule{0.025\textwidth}{1pt}}\put(-72,63){\scriptsize 10\% Full}
    \put(-85,52){\color{blue}\rule{0.010\textwidth}{1pt}}\put(-79,52){\color{blue}\rule{0.010\textwidth}{1pt}}\put(-72,50){\scriptsize 5\% Resonant}
    \put(-85,39){\color{red}\rule{0.010\textwidth}{1pt}}\put(-79,39){\color{red}\rule{0.010\textwidth}{1pt}}\put(-72,37){\scriptsize 10\% Resonant}
    \caption{The parton level distributions of $m_{tt}$ from resonant production only (dashed curves) and with interference effects (solid curves) with two choices of the scalar width: $\Gamma_{H/A}/m_{H/A} = 5\%$ (blue) and $\Gamma_{H/A}/m_{H/A} = 10\%$ (red), assuming $m_{H/A}=500$ GeV, $t_\beta = 1$ and $c_{\beta-\alpha} = 0$.}
    \label{fig:mtt_HH_HA}
\end{figure}
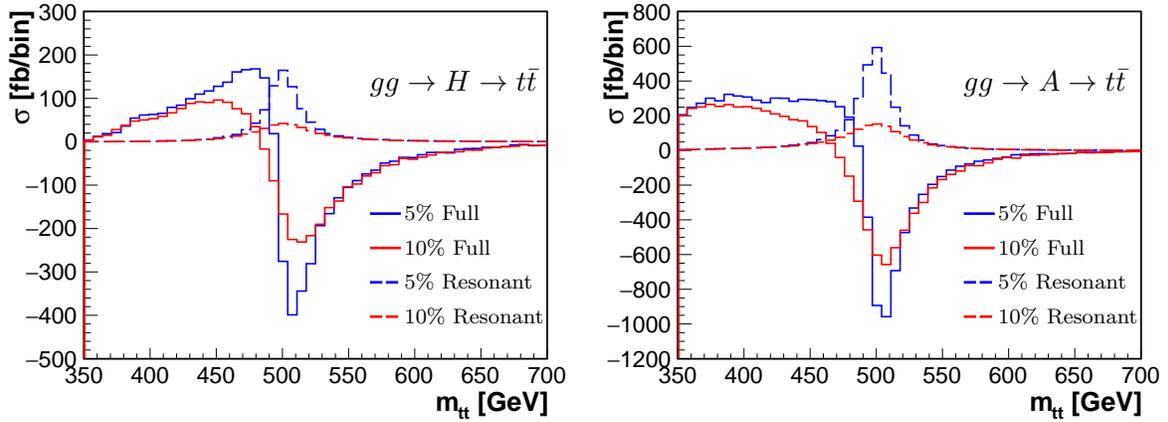

In~\autoref{fig:mtt_HH_HA}, we show the top pair invariant mass  distribution $m_{tt}$ at the parton level for the CP-even scalar $H$ (left panel) and CP-odd mode $A$ (right panel). The signal-only resonant profiles are shown by the dashed curves, while the full signal contributions, which account for interference effects, are shown in the solid curves. Interferences highly depend on the scalar width. Hence, in~\autoref{fig:mtt_HH_HA}, the distributions are shown for two different choices of  width: $\Gamma_{H/A}/m_{H/A}=5\%$ (blue) and $\Gamma_{H/A}/m_{H/A}=10\%$ (red). It is clear that the resonant shape from the signal-only sample changes into a bump-dip shape when accounting for the interference with the QCD top-quark pair production. Thus, it is necessary to include these effects to generate robust phenomenological modeling for the signal sample.

Current ATLAS and CMS analyses for resonant top-quark pair production account for the interferences with the SM $t\bar{t}$ production~\cite{ATLAS:2017snw,Sirunyan:2019wph}. In particular, the more recent CMS results are presented as upper limits on the coupling of the scalar to the top-quark pair as a function of the scalar mass and width. When we apply such constraints for each parameter point, the total width of the corresponding scalar is calculated. Then, the upper bound on the coupling of the scalar to top-quark pair is obtained by interpolating the results from the experimental results for the corresponding scalar mass. This upper bound is then compared with the actual coupling for this parameter point. This analysis can be translated into constraints on the  2HDM parameter space. Remarkably, it is found that the  top-quark pair searches $gg\to H/A\to t\bar t$ can cover a large portion of the parameter space sensitive SFOEWPT in the 2HDM~\cite{Goncalves:2021egx}.

\section*{Acknowledgements}
\label{sec:acknowledgements}
DG, AK, and YW thank the U.S.~Department of Energy for financial support, under grant number DE-SC 0016013. Some computing for this project was performed at the High-Performance Computing Center at Oklahoma State University, supported in part through the National Science Foundation grant OAC-1531128.

\bibliographystyle{JHEP}
\bibliography{references}

@article{Dorsch:2017nza,
    author = "Dorsch, G.C. and Huber, S.J. and Mimasu, K. and No, J.M.",
    title = "{The Higgs Vacuum Uplifted: Revisiting the Electroweak Phase Transition with a Second Higgs Doublet}",
    eprint = "1705.09186",
    archivePrefix = "arXiv",
    primaryClass = "hep-ph",
    reportNumber = "CP3-17-15, DESY-17-076, KCL-PH-TH-2017-27",
    doi = "10.1007/JHEP12(2017)086",
    journal = "JHEP",
    volume = "12",
    pages = "086",
    year = "2017"
}

@article{Bechtle:2020uwn,
    author = "Bechtle, Philip and Heinemeyer, Sven and Klingl, Tobias and Stefaniak, Tim and Weiglein, Georg and Wittbrodt, Jonas",
    title = "{HiggsSignals-2: Probing new physics with precision Higgs measurements in the LHC 13 TeV era}",
    eprint = "2012.09197",
    archivePrefix = "arXiv",
    primaryClass = "hep-ph",
    reportNumber = "BONN-TH-2020-09, DESY-20-228, DESY 20-228, IFT-UAM/CSIC-20-081, LU TP 20-53",
    doi = "10.1140/epjc/s10052-021-08942-y",
    journal = "Eur. Phys. J. C",
    volume = "81",
    number = "2",
    pages = "145",
    year = "2021"
}

@article{Bechtle:2020pkv,
    author = "Bechtle, Philip and Dercks, Daniel and Heinemeyer, Sven and Klingl, Tobias and Stefaniak, Tim and Weiglein, Georg and Wittbrodt, Jonas",
    title = "{HiggsBounds-5: Testing Higgs Sectors in the LHC 13 TeV Era}",
    eprint = "2006.06007",
    archivePrefix = "arXiv",
    primaryClass = "hep-ph",
    reportNumber = "BONN-TH-2020-03, DESY 20-093, DESY-20-093, IFT-UAM/CSIC-20-072, LU 20-27",
    doi = "10.1140/epjc/s10052-020-08557-9",
    month = "6",
    year = "2020"
}

@article{Muhlleitner:2020wwk,
    author = {M\"uhlleitner, Margarete and Sampaio, Marco O. P. and Santos, Rui and Wittbrodt, Jonas},
    title = "{ScannerS: Parameter Scans in Extended Scalar Sectors}",
    eprint = "2007.02985",
    archivePrefix = "arXiv",
    primaryClass = "hep-ph",
    reportNumber = "KA-TP-05-2020, LU TP 20-38",
    month = "7",
    year = "2020"
}

@article{Gunion:2002zf,
    author = "Gunion, John F. and Haber, Howard E.",
    title = "{The CP conserving two Higgs doublet model: The Approach to the decoupling limit}",
    eprint = "hep-ph/0207010",
    archivePrefix = "arXiv",
    reportNumber = "SCIPP-02-10",
    doi = "10.1103/PhysRevD.67.075019",
    journal = "Phys. Rev. D",
    volume = "67",
    pages = "075019",
    year = "2003"
}

@article{Han:2020lta,
    author = "Han, Tao and Li, Shuailong and Su, Shufang and Su, Wei and Wu, Yongcheng",
    title = "{Comparative Studies of 2HDMs under the Higgs Boson Precision Measurements}",
    eprint = "2008.05492",
    archivePrefix = "arXiv",
    primaryClass = "hep-ph",
    doi = "10.1007/JHEP01(2021)045",
    journal = "JHEP",
    volume = "01",
    pages = "045",
    year = "2021"
}

@article{PhysRevD.7.1888,
  title = {Radiative Corrections as the Origin of Spontaneous Symmetry Breaking},
  author = {Coleman, Sidney and Weinberg, Erick},
  journal = {Phys. Rev. D},
  volume = {7},
  issue = {6},
  pages = {1888--1910},
  numpages = {0},
  year = {1973},
  month = {Mar},
  publisher = {American Physical Society},
  doi = {10.1103/PhysRevD.7.1888},
  url = {https://link.aps.org/doi/10.1103/PhysRevD.7.1888}
}

@article{Basler:2016obg,
    author = "Basler, P. and Krause, M. and Muhlleitner, M. and Wittbrodt, J. and Wlotzka, A.",
    title = "{Strong First Order Electroweak Phase Transition in the CP-Conserving 2HDM Revisited}",
    eprint = "1612.04086",
    archivePrefix = "arXiv",
    primaryClass = "hep-ph",
    doi = "10.1007/JHEP02(2017)121",
    journal = "JHEP",
    volume = "02",
    pages = "121",
    year = "2017"
}

@article{Camargo-Molina:2016moz,
    author = "Camargo-Molina, Jos\'e Eliel and Morais, Ant\'onio P. and Pasechnik, Roman and Sampaio, Marco O. P. and Wess\'en, Jonas",
    title = "{All one-loop scalar vertices in the effective potential approach}",
    eprint = "1606.07069",
    archivePrefix = "arXiv",
    primaryClass = "hep-ph",
    doi = "10.1007/JHEP08(2016)073",
    journal = "JHEP",
    volume = "08",
    pages = "073",
    year = "2016"
}

@article{Arnold:1992rz,
    author = "Arnold, Peter Brockway and Espinosa, Olivier",
    title = "{The Effective potential and first order phase transitions: Beyond leading-order}",
    eprint = "hep-ph/9212235",
    archivePrefix = "arXiv",
    reportNumber = "UW-PT-92-18, USM-TH-60",
    doi = "10.1103/PhysRevD.47.3546",
    journal = "Phys. Rev. D",
    volume = "47",
    pages = "3546",
    year = "1993",
    note = "[Erratum: Phys.Rev.D 50, 6662 (1994)]"
}

@inproceedings{Quiros:1999jp,
    author = "Quiros, Mariano",
    title = "{Finite temperature field theory and phase transitions}",
    booktitle = "{ICTP Summer School in High-Energy Physics and Cosmology}",
    eprint = "hep-ph/9901312",
    archivePrefix = "arXiv",
    reportNumber = "IEM-FT-187-99",
    month = "1",
    year = "1999"
}

@article{Ramsey-Musolf:2019lsf,
    author = "Ramsey-Musolf, Michael J.",
    title = "{The electroweak phase transition: a collider target}",
    eprint = "1912.07189",
    archivePrefix = "arXiv",
    primaryClass = "hep-ph",
    reportNumber = "ACFI-T19-14",
    doi = "10.1007/JHEP09(2020)179",
    journal = "JHEP",
    volume = "09",
    pages = "179",
    year = "2020"
}

@article{EWPT-NMSSM,
    author = "Weicong, Huang and Zhaofeng, Kang and Jing, Shu and Peiwen, Wu and Jin, Min, Yang",
    title = "{New insights in the electroweak phase transition in the NMSSM}",
    eprint = "1405.1152",
    archivePrefix = "arXiv",
    primaryClass = "hep-ph",
    doi = "10.1103/PhysRevD.91.025006",
    journal = "Phys. Rev. D",
    volume = "91",
    issue = "2",
    pages = "025006",
    year = "2015"
}

@article{EWPT-Nature,
    author = "Harman, C.P and Huber, S.J",
    title = "{Does zero temperature decide on the nature of theelectroweak phase transition?}",
    eprint = "1512.05611",
    archivePrefix = "arXiv",
    primaryClass = "hep-ph",
    doi = "10.1103/PhysRevD.91.025006",
    journal = "Journal of High Energy Physics",
    volume = "2016",
    pages = "1-35",
    year = "2015"
}

@article{Sirunyan:2019wph,
    author = "Sirunyan, Albert M and others",
    collaboration = "CMS",
    title = "{Search for heavy Higgs bosons decaying to a top quark pair in proton-proton collisions at $\sqrt{s} =$ 13 TeV}",
    eprint = "1908.01115",
    archivePrefix = "arXiv",
    primaryClass = "hep-ex",
    reportNumber = "CMS-HIG-17-027, CERN-EP-2019-147",
    doi = "10.1007/JHEP04(2020)171",
    journal = "JHEP",
    volume = "04",
    pages = "171",
    year = "2020"
}

@article{Sirunyan:2019wrn,
    author = "Sirunyan, Albert M and others",
    collaboration = "CMS",
    title = "{Search for new neutral Higgs bosons through the H$\to$ ZA $\to \ell^{+}\ell^{-} \mathrm{b\bar{b}}$ process in pp collisions at $\sqrt{s} =$ 13 TeV}",
    eprint = "1911.03781",
    archivePrefix = "arXiv",
    primaryClass = "hep-ex",
    reportNumber = "CMS-HIG-18-012, CERN-EP-2019-254",
    doi = "10.1007/JHEP03(2020)055",
    journal = "JHEP",
    volume = "03",
    pages = "055",
    year = "2020"
}

@article{Wainwright:2011kj,
    author = "Wainwright, Carroll L.",
    title = "{CosmoTransitions: Computing Cosmological Phase Transition Temperatures and Bubble Profiles with Multiple Fields}",
    eprint = "1109.4189",
    archivePrefix = "arXiv",
    primaryClass = "hep-ph",
    doi = "10.1016/j.cpc.2012.04.004",
    journal = "Comput. Phys. Commun.",
    volume = "183",
    pages = "2006--2013",
    year = "2012"
}

@article{Grojean:2006bp,
      author         = "Grojean, Christophe and Servant, Geraldine",
      title          = "{Gravitational Waves from Phase Transitions at the
                        Electroweak Scale and Beyond}",
      journal        = "Phys. Rev.",
      volume         = "D75",
      year           = "2007",
      pages          = "043507",
      doi            = "10.1103/PhysRevD.75.043507",
      eprint         = "hep-ph/0607107",
      archivePrefix  = "arXiv",
      primaryClass   = "hep-ph",
      reportNumber   = "CERN-PH-TH-2006-125",
      SLACcitation   = "%%CITATION = HEP-PH/0607107;%%"
}

@article{Audley:2017drz,
      author         = "Audley, Heather and others",
      title          = "{Laser Interferometer Space Antenna}",
      collaboration  = "LISA",
      year           = "2017",
      eprint         = "1702.00786",
      archivePrefix  = "arXiv",
      primaryClass   = "astro-ph.IM",
      SLACcitation   = "%%CITATION = ARXIV:1702.00786;%%"
}

@article{Caprini:2015zlo,
      author         = "Caprini, Chiara and others",
      title          = "{Science with the space-based interferometer eLISA. II:
                        Gravitational waves from cosmological phase transitions}",
      journal        = "JCAP",
      volume         = "1604",
      year           = "2016",
      number         = "04",
      pages          = "001",
      doi            = "10.1088/1475-7516/2016/04/001",
      eprint         = "1512.06239",
      archivePrefix  = "arXiv",
      primaryClass   = "astro-ph.CO",
      reportNumber   = "DESY-15-246",
      SLACcitation   = "%%CITATION = ARXIV:1512.06239;%%"
}

@article{Aad:2020ncx,
    author = "Aad, Georges and others",
    collaboration = "ATLAS",
    title = "{Search for a heavy Higgs boson decaying into a Z boson and another heavy Higgs boson in the $\ell \ell bb$ and $\ell \ell WW$ final states in $pp$ collisions at $\sqrt{s}=13$~$\text {TeV}$ with the ATLAS detector}",
    eprint = "2011.05639",
    archivePrefix = "arXiv",
    primaryClass = "hep-ex",
    reportNumber = "CERN-EP-2020-191",
    doi = "10.1140/epjc/s10052-021-09117-5",
    journal = "Eur. Phys. J. C",
    volume = "81",
    number = "5",
    pages = "396",
    year = "2021"
}

@article{Moore:1998swa,
    author = "Moore, Guy D.",
    title = "{Measuring the broken phase sphaleron rate nonperturbatively}",
    eprint = "hep-ph/9805264",
    archivePrefix = "arXiv",
    reportNumber = "MCGILL-98-7",
    doi = "10.1103/PhysRevD.59.014503",
    journal = "Phys. Rev. D",
    volume = "59",
    pages = "014503",
    year = "1999"
}

@article{Linde:1980tt,
    author = "Linde, Andrei D.",
    title = "{Fate of the False Vacuum at Finite Temperature: Theory and Applications}",
    reportNumber = "LEBEDEV-80-92",
    doi = "10.1016/0370-2693(81)90281-1",
    journal = "Phys. Lett. B",
    volume = "100",
    pages = "37--40",
    year = "1981"
}

@article{Coleman:1977py,
    author = "Coleman, Sidney R.",
    title = "{The Fate of the False Vacuum. 1. Semiclassical Theory}",
    reportNumber = "HUTP-77-A004",
    doi = "10.1103/PhysRevD.16.1248",
    journal = "Phys. Rev. D",
    volume = "15",
    pages = "2929--2936",
    year = "1977",
    note = "[Erratum: Phys.Rev.D 16, 1248 (1977)]"
}

@article{Moreno:1998bq,
    author = "Moreno, J. M. and Quiros, M. and Seco, M.",
    title = "{Bubbles in the supersymmetric standard model}",
    eprint = "hep-ph/9801272",
    archivePrefix = "arXiv",
    reportNumber = "IEM-FT-168-98",
    doi = "10.1016/S0550-3213(98)00283-1",
    journal = "Nucl. Phys. B",
    volume = "526",
    pages = "489--500",
    year = "1998"
}

@article{Goncalves:2018qas,
    author = "Gon\c{c}alves, Dorival and Han, Tao and Kling, Felix and Plehn, Tilman and Takeuchi, Michihisa",
    title = "{Higgs boson pair production at future hadron colliders: From kinematics to dynamics}",
    eprint = "1802.04319",
    archivePrefix = "arXiv",
    primaryClass = "hep-ph",
    reportNumber = "PITT-PACC-1802, UCI-TR-2018-1, IPMU-18-0028",
    doi = "10.1103/PhysRevD.97.113004",
    journal = "Phys. Rev. D",
    volume = "97",
    number = "11",
    pages = "113004",
    year = "2018"
}

@article{Plehn:1996wb,
    author = "Plehn, T. and Spira, M. and Zerwas, P. M.",
    title = "{Pair production of neutral Higgs particles in gluon-gluon collisions}",
    eprint = "hep-ph/9603205",
    archivePrefix = "arXiv",
    reportNumber = "DESY-95-215",
    doi = "10.1016/0550-3213(96)00418-X",
    journal = "Nucl. Phys. B",
    volume = "479",
    pages = "46--64",
    year = "1996",
    note = "[Erratum: Nucl.Phys.B 531, 655--655 (1998)]"
}

@article{Barman:2020ulr,
    author = "Barman, Rahool Kumar and Englert, Christoph and Gon\c{c}alves, Dorival and Spannowsky, Michael",
    title = "{Di-Higgs resonance searches in weak boson fusion}",
    eprint = "2007.07295",
    archivePrefix = "arXiv",
    primaryClass = "hep-ph",
    reportNumber = "IPPP/20/28, OSU-HEP-20-08",
    doi = "10.1103/PhysRevD.102.055014",
    journal = "Phys. Rev. D",
    volume = "102",
    number = "5",
    pages = "055014",
    year = "2020"
}

@article{Dorsch:2014qja,
    author = "Dorsch, G. C. and Huber, S. J. and Mimasu, K. and No, J. M.",
    title = "{Echoes of the Electroweak Phase Transition: Discovering a second Higgs doublet through $A_0 \rightarrow ZH_0$}",
    eprint = "1405.5537",
    archivePrefix = "arXiv",
    primaryClass = "hep-ph",
    doi = "10.1103/PhysRevLett.113.211802",
    journal = "Phys. Rev. Lett.",
    volume = "113",
    number = "21",
    pages = "211802",
    year = "2014"
}

@article{Kajantie:1996mn,
    author = "Kajantie, K. and Laine, M. and Rummukainen, K. and Shaposhnikov, Mikhail E.",
    title = "{Is there a hot electroweak phase transition at m(H) larger or equal to m(W)?}",
    eprint = "hep-ph/9605288",
    archivePrefix = "arXiv",
    reportNumber = "CERN-TH-96-126, HD-THEP-96-15, IUHET-333",
    doi = "10.1103/PhysRevLett.77.2887",
    journal = "Phys. Rev. Lett.",
    volume = "77",
    pages = "2887--2890",
    year = "1996"
}

@article{Sakharov:1967dj,
    author = "Sakharov, A. D.",
    title = "{Violation of CP Invariance, C asymmetry, and baryon asymmetry of the universe}",
    doi = "10.1070/PU1991v034n05ABEH002497",
    journal = "Pisma Zh. Eksp. Teor. Fiz.",
    volume = "5",
    pages = "32--35",
    year = "1967"
}

@article{Morrissey:2012db,
    author = "Morrissey, David E. and Ramsey-Musolf, Michael J.",
    title = "{Electroweak baryogenesis}",
    eprint = "1206.2942",
    archivePrefix = "arXiv",
    primaryClass = "hep-ph",
    reportNumber = "NPAC-12-08",
    doi = "10.1088/1367-2630/14/12/125003",
    journal = "New J. Phys.",
    volume = "14",
    pages = "125003",
    year = "2012"
}

@article{Trodden:1998ym,
    author = "Trodden, Mark",
    title = "{Electroweak baryogenesis}",
    eprint = "hep-ph/9803479",
    archivePrefix = "arXiv",
    reportNumber = "CWRU-P6-98",
    doi = "10.1103/RevModPhys.71.1463",
    journal = "Rev. Mod. Phys.",
    volume = "71",
    pages = "1463--1500",
    year = "1999"
}

@article{Cohen:1993nk,
    author = "Cohen, Andrew G. and Kaplan, D. B. and Nelson, A. E.",
    title = "{Progress in electroweak baryogenesis}",
    eprint = "hep-ph/9302210",
    archivePrefix = "arXiv",
    reportNumber = "UCSD-PTH-93-02, BUHEP-93-4",
    doi = "10.1146/annurev.ns.43.120193.000331",
    journal = "Ann. Rev. Nucl. Part. Sci.",
    volume = "43",
    pages = "27--70",
    year = "1993"
}

@article{Carena:1996wj,
    author = "Carena, Marcela and Quiros, M. and Wagner, C. E. M.",
    title = "{Opening the window for electroweak baryogenesis}",
    eprint = "hep-ph/9603420",
    archivePrefix = "arXiv",
    reportNumber = "CERN-TH-96-30, IEM-FT-126-96",
    doi = "10.1016/0370-2693(96)00475-3",
    journal = "Phys. Lett. B",
    volume = "380",
    pages = "81--91",
    year = "1996"
}

@article{Eboli:1987dy,
      author         = "Eboli, Oscar J. P. and Marques, G. C. and Novaes, S. F.
                        and Natale, A. A.",
      title          = "{TWIN HIGGS BOSON PRODUCTION}",
      journal        = "Phys. Lett.",
      volume         = "B197",
      year           = "1987",
      pages          = "269-272",
      doi            = "10.1016/0370-2693(87)90381-9",
      reportNumber   = "MIT-CTP-1376A",
      SLACcitation   = "%%CITATION = PHLTA,B197,269;%%"
}

@article{No:2013wsa,
    author = "No, Jose M. and Ramsey-Musolf, Michael",
    title = "{Probing the Higgs Portal at the LHC Through Resonant di-Higgs Production}",
    eprint = "1310.6035",
    archivePrefix = "arXiv",
    primaryClass = "hep-ph",
    doi = "10.1103/PhysRevD.89.095031",
    journal = "Phys. Rev. D",
    volume = "89",
    number = "9",
    pages = "095031",
    year = "2014"
}

@article{Su:2020pjw,
    author = "Su, Wei and Williams, Anthony G. and Zhang, Mengchao",
    title = "{Strong first order electroweak phase transition in 2HDM confronting future Z \& Higgs factories}",
    eprint = "2011.04540",
    archivePrefix = "arXiv",
    primaryClass = "hep-ph",
    reportNumber = "ADP-20-31/T1141",
    doi = "10.1007/JHEP04(2021)219",
    journal = "JHEP",
    volume = "04",
    pages = "219",
    year = "2021"
}

@article{Branco:2011iw,
    author = "Branco, G. C. and Ferreira, P. M. and Lavoura, L. and Rebelo, M. N. and Sher, Marc and Silva, Joao P.",
    title = "{Theory and phenomenology of two-Higgs-doublet models}",
    eprint = "1106.0034",
    archivePrefix = "arXiv",
    primaryClass = "hep-ph",
    doi = "10.1016/j.physrep.2012.02.002",
    journal = "Phys. Rept.",
    volume = "516",
    pages = "1--102",
    year = "2012"
}

@article{Bernon:2017jgv,
    author = "Bernon, J\'er\'emy and Bian, Ligong and Jiang, Yun",
    title = "{A new insight into the phase transition in the early Universe with two Higgs doublets}",
    eprint = "1712.08430",
    archivePrefix = "arXiv",
    primaryClass = "hep-ph",
    doi = "10.1007/JHEP05(2018)151",
    journal = "JHEP",
    volume = "05",
    pages = "151",
    year = "2018"
}

@article{Aoki:2021oez,
    author = "Aoki, Mayumi and Komatsu, Takatoshi and Shibuya, Hiroto",
    title = "{Possibility of multi-step electroweak phase transition in the two Higgs doublet models}",
    eprint = "2106.03439",
    archivePrefix = "arXiv",
    primaryClass = "hep-ph",
    reportNumber = "KANAZAWA-21-08",
    month = "6",
    year = "2021"
}

@article{Andersen:2017ika,
    author = "Andersen, Jens O. and Gorda, Tyler and Helset, Andreas and Niemi, Lauri and Tenkanen, Tuomas V. I. and Tranberg, Anders and Vuorinen, Aleksi and Weir, David J.",
    title = "{Nonperturbative Analysis of the Electroweak Phase Transition in the Two Higgs Doublet Model}",
    eprint = "1711.09849",
    archivePrefix = "arXiv",
    primaryClass = "hep-ph",
    reportNumber = "HIP-2017-26/TH, HIP-2017-26-TH",
    doi = "10.1103/PhysRevLett.121.191802",
    journal = "Phys. Rev. Lett.",
    volume = "121",
    number = "19",
    pages = "191802",
    year = "2018"
}

@article{Dorsch:2013wja,
    author = "Dorsch, G. C. and Huber, S. J. and No, J. M.",
    title = "{A strong electroweak phase transition in the 2HDM after LHC8}",
    eprint = "1305.6610",
    archivePrefix = "arXiv",
    primaryClass = "hep-ph",
    doi = "10.1007/JHEP10(2013)029",
    journal = "JHEP",
    volume = "10",
    pages = "029",
    year = "2013"
}

@article{Dorsch:2016tab,
    author = "Dorsch, G. C. and Huber, S. J. and Mimasu, K. and No, J. M.",
    title = "{Hierarchical versus degenerate 2HDM: The LHC run 1 legacy at the onset of run 2}",
    eprint = "1601.04545",
    archivePrefix = "arXiv",
    primaryClass = "hep-ph",
    reportNumber = "DESY-15-240",
    doi = "10.1103/PhysRevD.93.115033",
    journal = "Phys. Rev. D",
    volume = "93",
    number = "11",
    pages = "115033",
    year = "2016"
}

@article{Davoudiasl:2021syn,
    author = "Davoudiasl, Hooman and Lewis, Ian M. and Sullivan, Matthew",
    title = "{Multi-TeV signals of baryogenesis in a Higgs troika model}",
    eprint = "2103.12089",
    archivePrefix = "arXiv",
    primaryClass = "hep-ph",
    doi = "10.1103/PhysRevD.104.015024",
    journal = "Phys. Rev. D",
    volume = "104",
    number = "1",
    pages = "015024",
    year = "2021"
}

@article{Goncalves:2021egx,
    author = "Gon\c{c}alves, Dorival and Kaladharan, Ajay and Wu, Yongcheng",
    title = "{Electroweak phase transition in the 2HDM: Collider and gravitational wave complementarity}",
    eprint = "2108.05356",
    archivePrefix = "arXiv",
    primaryClass = "hep-ph",
    doi = "10.1103/PhysRevD.105.095041",
    journal = "Phys. Rev. D",
    volume = "105",
    number = "9",
    pages = "095041",
    year = "2022"
}

@article{Dicus:1994bm,
    author = "Dicus, D. and Stange, A. and Willenbrock, S.",
    title = "{Higgs decay to top quarks at hadron colliders}",
    eprint = "hep-ph/9404359",
    archivePrefix = "arXiv",
    reportNumber = "CPP-94-18, BNL-60339, ILL-TH-94-9",
    doi = "10.1016/0370-2693(94)91017-0",
    journal = "Phys. Lett. B",
    volume = "333",
    pages = "126--131",
    year = "1994"
}

@article{Carena:2016npr,
    author = "Carena, Marcela and Liu, Zhen",
    title = "{Challenges and opportunities for heavy scalar searches in the $ t\overline{t} $ channel at the LHC}",
    eprint = "1608.07282",
    archivePrefix = "arXiv",
    primaryClass = "hep-ph",
    reportNumber = "FERMILAB-PUB-16-262-T",
    doi = "10.1007/JHEP11(2016)159",
    journal = "JHEP",
    volume = "11",
    pages = "159",
    year = "2016"
}

@article{Djouadi:2019cbm,
    author = "Djouadi, Abdelhak and Ellis, John and Popov, Andrey and Quevillon, J\'er\'emie",
    title = "{Interference effects in $ t\overline{t} $ production at the LHC as a window on new physics}",
    eprint = "1901.03417",
    archivePrefix = "arXiv",
    primaryClass = "hep-ph",
    reportNumber = "CERN-TH-2019-001, KCL-PH-TH/2018-05, LAPTH/001/19",
    doi = "10.1007/JHEP03(2019)119",
    journal = "JHEP",
    volume = "03",
    pages = "119",
    year = "2019"
}

@article{Alwall:2014hca,
    author = "Alwall, J. and Frederix, R. and Frixione, S. and Hirschi, V. and Maltoni, F. and Mattelaer, O. and Shao, H. -S. and Stelzer, T. and Torrielli, P. and Zaro, M.",
    title = "{The automated computation of tree-level and next-to-leading order differential cross sections, and their matching to parton shower simulations}",
    eprint = "1405.0301",
    archivePrefix = "arXiv",
    primaryClass = "hep-ph",
    reportNumber = "CERN-PH-TH-2014-064, CP3-14-18, LPN14-066, MCNET-14-09, ZU-TH-14-14",
    doi = "10.1007/JHEP07(2014)079",
    journal = "JHEP",
    volume = "07",
    pages = "079",
    year = "2014"
}

@article{Sjostrand:2014zea,
    author = {Sj\"ostrand, Torbj\"orn and Ask, Stefan and Christiansen, Jesper R. and Corke, Richard and Desai, Nishita and Ilten, Philip and Mrenna, Stephen and Prestel, Stefan and Rasmussen, Christine O. and Skands, Peter Z.},
    title = "{An introduction to PYTHIA 8.2}",
    eprint = "1410.3012",
    archivePrefix = "arXiv",
    primaryClass = "hep-ph",
    reportNumber = "LU-TP-14-36, MCNET-14-22, CERN-PH-TH-2014-190, FERMILAB-PUB-14-316-CD, DESY-14-178, SLAC-PUB-16122",
    doi = "10.1016/j.cpc.2015.01.024",
    journal = "Comput. Phys. Commun.",
    volume = "191",
    pages = "159--177",
    year = "2015"
}

@article{deFavereau:2013fsa,
    author = "de Favereau, J. and Delaere, C. and Demin, P. and Giammanco, A. and Lema\^\i{}tre, V. and Mertens, A. and Selvaggi, M.",
    collaboration = "DELPHES 3",
    title = "{DELPHES 3, A modular framework for fast simulation of a generic collider experiment}",
    eprint = "1307.6346",
    archivePrefix = "arXiv",
    primaryClass = "hep-ex",
    doi = "10.1007/JHEP02(2014)057",
    journal = "JHEP",
    volume = "02",
    pages = "057",
    year = "2014"
}

@article{Cacciari:2011ma,
    author = "Cacciari, Matteo and Salam, Gavin P. and Soyez, Gregory",
    title = "{FastJet User Manual}",
    eprint = "1111.6097",
    archivePrefix = "arXiv",
    primaryClass = "hep-ph",
    reportNumber = "CERN-PH-TH-2011-297",
    doi = "10.1140/epjc/s10052-012-1896-2",
    journal = "Eur. Phys. J. C",
    volume = "72",
    pages = "1896",
    year = "2012"
}

@article{Alloul:2013bka,
    author = "Alloul, Adam and Christensen, Neil D. and Degrande, C\'eline and Duhr, Claude and Fuks, Benjamin",
    title = "{FeynRules  2.0 - A complete toolbox for tree-level phenomenology}",
    eprint = "1310.1921",
    archivePrefix = "arXiv",
    primaryClass = "hep-ph",
    reportNumber = "CERN-PH-TH-2013-239, MCNET-13-14, IPPP-13-71, DCPT-13-142, PITT-PACC-1308",
    doi = "10.1016/j.cpc.2014.04.012",
    journal = "Comput. Phys. Commun.",
    volume = "185",
    pages = "2250--2300",
    year = "2014"
}

@article{Degrande:2014vpa,
    author = "Degrande, Celine",
    title = "{Automatic evaluation of UV and R2 terms for beyond the Standard Model Lagrangians: a proof-of-principle}",
    eprint = "1406.3030",
    archivePrefix = "arXiv",
    primaryClass = "hep-ph",
    reportNumber = "IPPP-14-58, DCPT-14-116",
    doi = "10.1016/j.cpc.2015.08.015",
    journal = "Comput. Phys. Commun.",
    volume = "197",
    pages = "239--262",
    year = "2015"
}

@article{Gaemers:1984sj,
    author = "Gaemers, K. J. F. and Hoogeveen, F.",
    title = "{Higgs Production and Decay Into Heavy Flavors With the Gluon Fusion Mechanism}",
    reportNumber = "ITFA-84-6",
    doi = "10.1016/0370-2693(84)91711-8",
    journal = "Phys. Lett. B",
    volume = "146",
    pages = "347--349",
    year = "1984"
}

@article{Bernreuther:1997gs,
    author = "Bernreuther, W. and Flesch, M. and Haberl, P.",
    title = "{Signatures of Higgs bosons in the top quark decay channel at hadron colliders}",
    eprint = "hep-ph/9709284",
    archivePrefix = "arXiv",
    reportNumber = "PITHA-97-34",
    doi = "10.1103/PhysRevD.58.114031",
    journal = "Phys. Rev. D",
    volume = "58",
    pages = "114031",
    year = "1998"
}

@article{Barger:2006hm,
    author = "Barger, Vernon and Han, Tao and Walker, Devin G. E.",
    title = "{Top Quark Pairs at High Invariant Mass: A Model-Independent Discriminator of New Physics at the LHC}",
    eprint = "hep-ph/0612016",
    archivePrefix = "arXiv",
    reportNumber = "MADPH-06-1475",
    doi = "10.1103/PhysRevLett.100.031801",
    journal = "Phys. Rev. Lett.",
    volume = "100",
    pages = "031801",
    year = "2008"
}

@article{Frederix:2007gi,
    author = "Frederix, Rikkert and Maltoni, Fabio",
    title = "{Top pair invariant mass distribution: A Window on new physics}",
    eprint = "0712.2355",
    archivePrefix = "arXiv",
    primaryClass = "hep-ph",
    reportNumber = "CP3-07-29",
    doi = "10.1088/1126-6708/2009/01/047",
    journal = "JHEP",
    volume = "01",
    pages = "047",
    year = "2009"
}

@article{Barcelo:2010bm,
    author = "Barcelo, Roberto and Masip, Manuel",
    title = "{Extra Higgs bosons in $t\bar{t}$ production at the LHC}",
    eprint = "1001.5456",
    archivePrefix = "arXiv",
    primaryClass = "hep-ph",
    doi = "10.1103/PhysRevD.81.075019",
    journal = "Phys. Rev. D",
    volume = "81",
    pages = "075019",
    year = "2010"
}

@article{Figy:2011yu,
    author = "Figy, Terrance and Zwicky, Roman",
    title = "{The other Higgses, at resonance, in the Lee-Wick extension of the Standard Model}",
    eprint = "1108.3765",
    archivePrefix = "arXiv",
    primaryClass = "hep-ph",
    reportNumber = "CERN-PH-TH-2011-184, CP-3-2011-25",
    doi = "10.1007/JHEP10(2011)145",
    journal = "JHEP",
    volume = "10",
    pages = "145",
    year = "2011"
}

@article{Barger:2011pu,
    author = "Barger, Vernon and Keung, Wai-Yee and Yencho, Brian",
    title = "{Azimuthal Correlations in Top Pair Decays and The Effects of New Heavy Scalars}",
    eprint = "1112.5173",
    archivePrefix = "arXiv",
    primaryClass = "hep-ph",
    doi = "10.1103/PhysRevD.85.034016",
    journal = "Phys. Rev. D",
    volume = "85",
    pages = "034016",
    year = "2012"
}

@article{Moretti:2012mq,
    author = "Moretti, S. and Ross, D. A.",
    title = "{On the top-antitop invariant mass spectrum at the LHC from a Higgs boson signal perspective}",
    eprint = "1203.3746",
    archivePrefix = "arXiv",
    primaryClass = "hep-ph",
    reportNumber = "SHEP-12-04",
    doi = "10.1016/j.physletb.2012.04.074",
    journal = "Phys. Lett. B",
    volume = "712",
    pages = "245--249",
    year = "2012"
}

@article{Craig:2015jba,
    author = "Craig, Nathaniel and D'Eramo, Francesco and Draper, Patrick and Thomas, Scott and Zhang, Hao",
    title = "{The Hunt for the Rest of the Higgs Bosons}",
    eprint = "1504.04630",
    archivePrefix = "arXiv",
    primaryClass = "hep-ph",
    doi = "10.1007/JHEP06(2015)137",
    journal = "JHEP",
    volume = "06",
    pages = "137",
    year = "2015"
}

@article{Bernreuther:2015fts,
    author = "Bernreuther, W. and Galler, P. and Mellein, C. and Si, Z. G. and Uwer, P.",
    title = "{Production of heavy Higgs bosons and decay into top quarks at the LHC}",
    eprint = "1511.05584",
    archivePrefix = "arXiv",
    primaryClass = "hep-ph",
    reportNumber = "HU-EP-15-52, TTK-15-33",
    doi = "10.1103/PhysRevD.93.034032",
    journal = "Phys. Rev. D",
    volume = "93",
    number = "3",
    pages = "034032",
    year = "2016"
}

@article{Gori:2016zto,
    author = "Gori, Stefania and Kim, Ian-Woo and Shah, Nausheen R. and Zurek, Kathryn M.",
    title = "{Closing the Wedge: Search Strategies for Extended Higgs Sectors with Heavy Flavor Final States}",
    eprint = "1602.02782",
    archivePrefix = "arXiv",
    primaryClass = "hep-ph",
    reportNumber = "WSU-HEP-1508",
    doi = "10.1103/PhysRevD.93.075038",
    journal = "Phys. Rev. D",
    volume = "93",
    number = "7",
    pages = "075038",
    year = "2016"
}

@article{Djouadi:2016ack,
    author = "Djouadi, Abdelhak and Ellis, John and Quevillon, J\'er\'emie",
    title = "{Interference effects in the decays of spin-zero resonances into $\gamma \gamma$ and $ t\overline{t} $}",
    eprint = "1605.00542",
    archivePrefix = "arXiv",
    primaryClass = "hep-ph",
    reportNumber = "CERN-TH-2016-095, KCL-PH-TH-2016-20, LPT-ORSAY-16-39, LPT-ORSAY--16--39",
    doi = "10.1007/JHEP07(2016)105",
    journal = "JHEP",
    volume = "07",
    pages = "105",
    year = "2016"
}

@article{Hespel:2016qaf,
    author = "Hespel, B. and Maltoni, F. and Vryonidou, E.",
    title = "{Signal background interference effects in heavy scalar production and decay to a top-anti-top pair}",
    eprint = "1606.04149",
    archivePrefix = "arXiv",
    primaryClass = "hep-ph",
    reportNumber = "CP3-16-30, MCNET-16-19",
    doi = "10.1007/JHEP10(2016)016",
    journal = "JHEP",
    volume = "10",
    pages = "016",
    year = "2016"
}

@article{Czakon:2016vfr,
    author = "Czakon, Michal and Heymes, David and Mitov, Alexander",
    title = "{Bump hunting in LHC $t\bar t$ events}",
    eprint = "1608.00765",
    archivePrefix = "arXiv",
    primaryClass = "hep-ph",
    reportNumber = "CAVENDISH-HEP-16-14, TTK-16-33",
    doi = "10.1103/PhysRevD.94.114033",
    journal = "Phys. Rev. D",
    volume = "94",
    number = "11",
    pages = "114033",
    year = "2016"
}

@article{Coimbra:2013qq,
    author = "Coimbra, Rita and Sampaio, Marco O. P. and Santos, Rui",
    title = "{ScannerS: Constraining the phase diagram of a complex scalar singlet at the LHC}",
    eprint = "1301.2599",
    archivePrefix = "arXiv",
    primaryClass = "hep-ph",
    doi = "10.1140/epjc/s10052-013-2428-4",
    journal = "Eur. Phys. J. C",
    volume = "73",
    pages = "2428",
    year = "2013"
}

@article{Lee:1977eg,
    author = "Lee, Benjamin W. and Quigg, C. and Thacker, H. B.",
    title = "{Weak Interactions at Very High-Energies: The Role of the Higgs Boson Mass}",
    reportNumber = "FERMILAB-PUB-77-30-THY, FERMILAB-PUB-77-030-T",
    doi = "10.1103/PhysRevD.16.1519",
    journal = "Phys. Rev. D",
    volume = "16",
    pages = "1519",
    year = "1977"
}

@article{Kanemura:1993hm,
    author = "Kanemura, Shinya and Kubota, Takahiro and Takasugi, Eiichi",
    title = "{Lee-Quigg-Thacker bounds for Higgs boson masses in a two doublet model}",
    eprint = "hep-ph/9303263",
    archivePrefix = "arXiv",
    reportNumber = "OS-GE-32-93",
    doi = "10.1016/0370-2693(93)91205-2",
    journal = "Phys. Lett. B",
    volume = "313",
    pages = "155--160",
    year = "1993"
}

@article{Ginzburg:2005dt,
    author = "Ginzburg, I. F. and Ivanov, I. P.",
    title = "{Tree-level unitarity constraints in the most general 2HDM}",
    eprint = "hep-ph/0508020",
    archivePrefix = "arXiv",
    doi = "10.1103/PhysRevD.72.115010",
    journal = "Phys. Rev. D",
    volume = "72",
    pages = "115010",
    year = "2005"
}

@article{Ivanov:2018jmz,
    author = {Ivanov, Igor P. and K\"opke, Marcel and M\"uhlleitner, Margarete},
    title = "{Algorithmic Boundedness-From-Below Conditions for Generic Scalar Potentials}",
    eprint = "1802.07976",
    archivePrefix = "arXiv",
    primaryClass = "hep-ph",
    doi = "10.1140/epjc/s10052-018-5893-y",
    journal = "Eur. Phys. J. C",
    volume = "78",
    number = "5",
    pages = "413",
    year = "2018"
}

@article{Hollik:2018wrr,
    author = "Hollik, Wolfgang G. and Weiglein, Georg and Wittbrodt, Jonas",
    title = "{Impact of Vacuum Stability Constraints on the Phenomenology of Supersymmetric Models}",
    eprint = "1812.04644",
    archivePrefix = "arXiv",
    primaryClass = "hep-ph",
    reportNumber = "DESY 18-148, DESY-18-148, TTP 18-036",
    doi = "10.1007/JHEP03(2019)109",
    journal = "JHEP",
    volume = "03",
    pages = "109",
    year = "2019"
}

@article{Ferreira:2019iqb,
    author = {Ferreira, P. M. and M\"uhlleitner, Margarete and Santos, Rui and Weiglein, Georg and Wittbrodt, Jonas},
    title = "{Vacuum Instabilities in the N2HDM}",
    eprint = "1905.10234",
    archivePrefix = "arXiv",
    primaryClass = "hep-ph",
    reportNumber = "DESY-19-085, DESY 19-085, KA-TP-07-2019",
    doi = "10.1007/JHEP09(2019)006",
    journal = "JHEP",
    volume = "09",
    pages = "006",
    year = "2019"
}

@article{ATLAS:2017snw,
    author = "Aaboud, Morad and others",
    collaboration = "ATLAS",
    title = "{Search for Heavy Higgs Bosons $A/H$ Decaying to a Top Quark Pair in $pp$ Collisions at $\sqrt{s}=8\text{ }\text{ }\mathrm{TeV}$ with the ATLAS Detector}",
    eprint = "1707.06025",
    archivePrefix = "arXiv",
    primaryClass = "hep-ex",
    reportNumber = "CERN-EP-2017-134",
    doi = "10.1103/PhysRevLett.119.191803",
    journal = "Phys. Rev. Lett.",
    volume = "119",
    number = "19",
    pages = "191803",
    year = "2017"
}

@article{Kainulainen:2019kyp,
    author = "Kainulainen, Kimmo and Keus, Venus and Niemi, Lauri and Rummukainen, Kari and Tenkanen, Tuomas V. I. and Vaskonen, Ville",
    title = "{On the validity of perturbative studies of the electroweak phase transition in the Two Higgs Doublet model}",
    eprint = "1904.01329",
    archivePrefix = "arXiv",
    primaryClass = "hep-ph",
    doi = "10.1007/JHEP06(2019)075",
    journal = "JHEP",
    volume = "06",
    pages = "075",
    year = "2019"
}

@article{Biekotter:2021ysx,
    author = {Biek\"otter, Thomas and Heinemeyer, Sven and No, Jos\'e Miguel and Olea, Mar\'\i{}a Olalla and Weiglein, Georg},
    title = "{Fate of electroweak symmetry in the early Universe: Non-restoration and trapped vacua in the N2HDM}",
    eprint = "2103.12707",
    archivePrefix = "arXiv",
    primaryClass = "hep-ph",
    reportNumber = "DESY-21-034, IFT-UAM/CSIC-21-018",
    doi = "10.1088/1475-7516/2021/06/018",
    journal = "JCAP",
    volume = "06",
    pages = "018",
    year = "2021"
}

@article{Enomoto:2021dkl,
    author = "Enomoto, Kazuki and Kanemura, Shinya and Mura, Yushi",
    title = "{Electroweak baryogenesis in aligned two Higgs doublet models}",
    eprint = "2111.13079",
    archivePrefix = "arXiv",
    primaryClass = "hep-ph",
    reportNumber = "OU-HET-1114",
    doi = "10.1007/JHEP01(2022)104",
    journal = "JHEP",
    volume = "01",
    pages = "104",
    year = "2022"
}

@article{Anisha:2022hgv,
    author = {Anisha and Biermann, Lisa and Englert, Christoph and M\"uhlleitner, Margarete},
    title = "{Two Higgs doublets, Effective Interactions and a Strong First-Order Electroweak Phase Transition}",
    eprint = "2204.06966",
    archivePrefix = "arXiv",
    primaryClass = "hep-ph",
    month = "4",
    year = "2022"
}

@article{Goncalves:2015mfa,
    author = {Goncalves, Dorival and Krauss, Frank and Kuttimalai, Silvan and Maierh\"ofer, Philipp},
    title = "{Higgs-Strahlung: Merging the NLO Drell-Yan and Loop-Induced 0+1 jet Multiplicities}",
    eprint = "1509.01597",
    archivePrefix = "arXiv",
    primaryClass = "hep-ph",
    reportNumber = "IPPP-15-57, DCPT-15-114",
    doi = "10.1103/PhysRevD.92.073006",
    journal = "Phys. Rev. D",
    volume = "92",
    number = "7",
    pages = "073006",
    year = "2015"
}

@article{Atkinson:2022pcn,
    author = "Atkinson, Oliver and Black, Matthew and Englert, Christoph and Lenz, Alexander and Rusov, Aleksey and Wynne, James",
    title = "{The Flavourful Present and Future of 2HDMs at the Collider Energy Frontier}",
    eprint = "2202.08807",
    archivePrefix = "arXiv",
    primaryClass = "hep-ph",
    reportNumber = "SI-HEP-2022-03",
    month = "2",
    year = "2022"
}

@article{Li:2020hao,
    author = "Li, Shuailong and Song, Huayang and Su, Shufang",
    title = "{Probing Exotic Charged Higgs Decays in the Type-II 2HDM through Top Rich Signal at a Future 100 TeV pp Collider}",
    eprint = "2005.00576",
    archivePrefix = "arXiv",
    primaryClass = "hep-ph",
    doi = "10.1007/JHEP11(2020)105",
    journal = "JHEP",
    volume = "11",
    pages = "105",
    year = "2020"
}

@article{Kling:2020hmi,
    author = "Kling, Felix and Su, Shufang and Su, Wei",
    title = "{2HDM Neutral Scalars under the LHC}",
    eprint = "2004.04172",
    archivePrefix = "arXiv",
    primaryClass = "hep-ph",
    doi = "10.1007/JHEP06(2020)163",
    journal = "JHEP",
    volume = "06",
    pages = "163",
    year = "2020"
}

\end{document}